\documentclass[traditabstract]{aa} % for the abstract without structuration 
\usepackage{graphicx}
\usepackage{txfonts}
\usepackage{natbib}
\bibpunct{(}{)}{;}{a}{}{,}
\newcommand{\halpha}{H$\alpha$}
\newcommand{\hbeta}{H$\beta$}

\newcommand{\msun}{M$_\odot$}
\newcommand{\mi}{$\mu$m}
\newcommand{\kms}{km~s$^{-1}$}
\newcommand{\htwo} {\hbox{H$_2$}}

\newcommand{\mmol} {\hbox{$M_{{\rm mol}}$}}
\newcommand{\mgas} {\hbox{$M_{{\rm gas}}$}}

\newcommand{\mhi}   {\hbox{$M_{\rm HI}$}}

\newcommand{\matom}   {\hbox{$M_{\rm atom}$}}

\newcommand{\sigmagas}{$\Sigma_{\rm gas}$}
\newcommand{\sigmamol}{$\Sigma_{\rm mol}$}

\newcommand{\sigmaatom}{$\Sigma_{\rm atom}$}
\newcommand{\sigmahtwo}{$\Sigma_{\rm H2}$}
\newcommand{\sigmasfr}{$\Sigma_{\rm SFR}$}

\newcommand{\spitzer}{{\it Spitzer}}
\newcommand{\tauorb}{$\tau_{\rm orb}$}
\newcommand{\taudep}{$\tau_{\rm dep}$}
\newcommand{\tauff}{$\tau_{\rm ff}$}

\newcommand{\xco}{$X_{\rm CO}$}
\begin{document}
   \title{Molecular gas and star formation in the Tidal Dwarf Galaxy \object{VCC~2062}}

%   \subtitle{I. Overviewing the $\kappa$-mechanism}

   \author{U. Lisenfeld\inst{1}
          \and
          J. Braine\inst{2}
           \and
          P.A. Duc\inst{3}
          \and
          M. Boquien\inst{4,5}
         \and
          E. Brinks\inst{6}
          \and
          F. Bournaud\inst{3}
          \and
           F. Lelli\inst{7}
           \and
          V. Charmandaris\inst{8,9}
          }
          
   \institute{Departamento de F\'isica Te\'orica y del Cosmos, Universidad de Granada, Spain and Instituto Carlos I de F\'isica T\'eorica y Computacional, 
   Facultad de Ciencias, 18071 Granada, Spain\\             
    \email{ute@ugr.es}
    \and
        Observatoire de Bordeaux, UMR 5804, CNRS/INSU, B.P. 89, F-33270 Floirac, France
  \and
    Laboratoire AIM, CEA/DSM - CNRS - Universit\'e Paris Diderot, DAPNIA/Service dÕAstrophysique, CEA-Saclay, F-91191
Gif-sur-Yvette Cedex, France
    \and
    Institute of Astronomy, University of Cambridge, Madingley Road, Cambridge, CB3 0HA, UK
    \and
    Unidad de Astronom\'\i a, Fac. Cs. B\'asicas, Universidad de Antofagasta, Avda. U. 
de Antofagasta 02800, Antofagasta, Chile
    \and
    Centre for Astrophysics Research, University of Hertfordshire, Hatfield AL10 9AB, UK
    \and
    Astronomy Department, Case Western Reserve University, 10900 Euclid Avenue, Cleveland, OH 44106
    \and
 Institute for Astronomy, Astrophysics, Space Applications and Remote Sensing, National Observatory of Athens, GR-15236, Penteli, Greece
 \and
 University of Crete, Department of Physics, Heraklion GR-71003, Greece
}
   \date{ }

% \abstract{}{}{}{}{} 
% 5 {} token are mandatory
 
  \abstract
  {
The physical mechanisms driving star formation (SF) in galaxies are still not fully understood. Tidal dwarf galaxies (TDGs), made of gas ejected during galaxy interactions, seem to be devoid of dark matter and have a near-solar metallicity.  The latter 
makes it possible to study molecular gas and its link to SF using standard tracers (CO, dust) in a peculiar environment.  
We present a detailed study of a nearby TDG in the Virgo Cluster, VCC 2062, using new high-resolution CO(1--0)\thanks{
The reduced data cubes are  available 
at the CDS via anonymous ftp to cdsarc.u-strasbg.fr (130.79.128.5)
or via http://cdsweb.u-strasbg.fr/cgi-bin/qcat?J/A+A/}
data from the Plateau de Bure, deep optical imaging from the Next Generation Virgo Cluster Survey (NGVS), and complementary multiwavelength data.
% (HI, \halpha, UV, 8 and 24~\mi).
%VCC~2062 is a nearby
%TDG and has a large set of ancillary data but was lacking high-resolution CO data.  
Until now, there was some doubt whether VCC~2062 was a true TDG, but the new deep optical images from the NGVS reveal a stellar bridge between VCC~2062 and its parent galaxy, NGC~4694, which is clear proof of its tidal origin. 
Several high-resolution tracers (\halpha, UV, 8~\mi, and 24~\mi) of the star formation rate (SFR) are compared to the molecular gas distribution as traced by  the CO(1--0). %data with the Plateau de Bure Interferometer at an angular resolution of  $\sim$ 4~\arcsec.  
Coupled with the SFR tracers, the NGVS data are used with the CIGALE code to model the stellar populations throughout VCC~2062,
yielding a declining SFR in the recent past, consistent with the low \halpha/UV ratio, and a high burst strength. 
HI emission covers VCC~2062, whereas the CO is concentrated near the HI maxima.  The CO peaks correspond to two very distinct regions: one with moderate SF to the NE and one with only slightly weaker CO emission but with nearly no SF.  Even where SF is clearly present, the SFR is below the value expected from the surface density of the molecular 
and the total gas as
compared to spiral galaxies and other TDGs.
% and also lower than what is expected from the gas surface density to free-fall time ratio, another SFR predictor.  
%
%%The discrepancy depends on the region on the spatial scale considered, being smallest for regiones center
%Given the strength of the CO emission, VCC~2062 has a very low SFR.
%for any SFR measure or predictor.  
%The stellar surface brightness is also very low 
%and the modelling with CIGALE yields a declining SFR in the recent past, consistent with the low \halpha/UV ratio, and a high burst strength. 
 After discussing different possible explanations, we conclude that the low surface brightness is a crucial parameter to understand the low SFR.
}


 \titlerunning{Molecular gas and star formation in the Tidal Dwarf Galaxy \object{VCC~2062}}
\authorrunning{ U. Lisenfeld,  et al.}

 \maketitle
%
%________________________________________________________________

\section{Introduction}

Tidal dwarf galaxies (TDGs) are self-gravitating entities formed from the tidal
debris of interacting galaxies \citep[see][and references therein]{duc12}. Their stellar and star formation (SF) properties are similar to those of dwarf
irregulars or blue compact dwarfs, with the noticeable exception that their
metallicity is higher (about 1/3 solar to 1/2 solar) because they are 
recycled objects, inheriting the metallicity of the outer regions of
their parent galaxies. Another relevant difference compared to non-tidal
dwarfs (called ``classical" dwarf galaxies in the following) is their expected lack of non-baryonic dark matter (DM) 
 if the standard Lambda cold dark matter ($\Lambda$CDM) cosmological
scenario is correct and  DM is distributed in a large, pressure-supported  halo. Since a typical tidal
interaction between two galaxies mostly expels material from the disk, 
it is predicted that only a fraction of DM resides in TDGs.
 This characteristic
is the strongest difference between TDGs and classical dwarfs,
but it is difficult to test because  the dynamical and the visible masses both
need to be measured \citep[see][for a review]{duc12}. A rough estimate of the dynamical mass from the CO(1-0) 
linewidth gave  a low dark-to-visible mass ratio of the order $\sim$ 1 in 8 TDGs
\citep{braine01}, considerably lower than the value typically found in classical
dwarfs \citep[baryon fraction of $\sim$ 30\%, e.g.][]{lelli14}.
In order to better determine the dynamical mass,
high-quality, high-resolution spectral data are necessary to spatially resolve the gas kinematics.
This has been done, so far,  in
three systems based on high-resolution HI data: \object{NGC~5291}, containing three TDGs \citep[][revised in Lelli et al. 2015]{bournaud07};
\object{NGC~7252} containing two objects \citep{lelli15}; and VCC~2062  \citep{lelli15}.  In all six TDGs, assuming
dynamical equilibrium, the dark-to-visible mass ratio
is consistent with 1 giving strong support to the tidal origin of the objects.

A major challenge in astrophysics is understanding the process of SF  in
galaxies and in particular how the star formation rate (SFR) is quantitatively related to the neutral gas
(the SF law). 
%A common way to study the SF law is to compare the surface density
%of the SFR to the surface density of the gas mass, the Kennicutt-Schmidt law (REF).
%The SFR can be compared to the total gas (atomic and molecular) or to the molecular gas.
%Studies of nearby galaxies has shown that the relation is tighter with the molecular gas
%than with the total gas
%( M51, Kennicutt 2007, M33, Heyer 2004, nearby
%spiral galaxies, Leroy, Bigiel) which is understandable because stars form from molecular gas.
%
A useful parameter for this kind of study is the star formation efficiency (SFE), which we define in this paper as the
SFR per molecular gas mass, SFE = SFR/\mmol\footnote{
Whenever we  refer to the SFE with respect to a different gas phase, we
state this explicitly, for example SFE(gas) = SFR/\mgas, the SFE per total neutral gas.
},
the inverse of the molecular gas consumption time, \taudep.
This parameter is relatively constant in nearby spiral galaxies 
with a median value of log(SFE) = -9.23 \citep{bigiel11}, suggesting that in these objects the SFR is simply determined by the
amount of available molecular gas.
Starburst  galaxies, however, form their
stars more efficiently with a SFE of about a factor  of four to 10 higher \citep{genzel10,daddi10}.
The  SFE at the dim end of the luminosity range is much more difficult to measure  because molecular gas is hard to detect and quantify.
In dwarf galaxies, the low metallicity makes CO a poor tracer of the 
molecular gas  \citep[see][for a review]{bolatto13}. 
This is partly due to a low molecular gas fraction, and partly due to a low
metallicity since the CO-to-H$_2$ conversion
factor, \xco , depends very sensitively on this latter parameter and also on the radiation field, making a quantitative measurement of
the molecular gas mass difficult even if CO is detected. %, so that the SFE could only be measured in a few objects. 
Only in nearby galaxies with high
sensitivity and  high-resolution CO data, allowing the  study of individual Giant Molecular Clouds (GMCs),
and with supplementary
infrared (IR) dust emission data, an estimate of the molecular gas mass is possible without the need to know
the CO-to-H$_2$ conversion factor.
In this way, 
in \object{M33}  (metallicity 0.5 solar) and \object{NGC 6822} (metallcity 0.3 solar)  increases
of the SFE of a factor 2-4 \citep[M33,][]{gratier10b}, and 5-10 \citep[NGC 6822,][]{gratier10a}   have been found.
A possible reason for this high SFE might be stronger stellar winds at low metallicities which
clear out more efficiently the surroundings of newly formed stars from molecular gas \citep{dib11}.
The lowest metallicity galaxies with CO detections are WLM and the SMC   (10-15\% of solar metallicity).
In the two molecular clouds detected in WLM, the SFE, at a scale of $\sim$  100 pc is very similar to
spiral galaxies, but a factor of 10 lower than in the Orion Nebula \citep{elmegreen13}.

TDGs possess several differences compared to 
other galaxies and in particular classical dwarfs 
which make the study of their SF law highly interesting.
First of all, due to 
the close-to-solar metallicity of TDGs the CO emission is a good tracer of the molecular
gas and allows a reliable measure of the molecular gas mass. 
Thus, TDGs allow us to directly study the SF law based on the molecular gas which is 
difficult  in other dwarf galaxies.
The preliminary  analysis of a small sample of TDGs with CO data and measurements of their SFR showed that 
these objects seem to follow the same relation as spiral galaxies \citep{lisenfeld02b}, with the exception of VCC~2062,  which showed a
low SFE. In order to study this low SFE in more detail, 
%However, the poor angular resolution of the existing CO data introduced considerable uncertainty and prompted  the need
high angular resolution
interferometric data are needed
\citep{lisenfeld09}.

A second difference is the low DM content. Even though DM does not have a direct relation to  SF, there are
secondary effects, for example on the rotation curve, which has consequences  for dynamically driven SF thresholds  \citep[e.g.][]{toomre64,hunter98}
or SF laws that are based on the orbital timescales \citep{kennicutt98}. 
%However,
%the low DM content does not seem to have fatal consequences for the stability of TDGs.
\citet{ploeckinger14,ploeckinger15} simulated the evolution of SF and chemistry  in TDGs and concluded that 
TDGs can survive starburst episodes without being disrupted in spite of their low DM content.  In their models,
even after an initial starburst,  the SF becomes self-regulated and continuous after a dynamical time.
Finally, in TDGs the environmental conditions are different among other things because of the presence of tidal forces.
The simulations of \citet{ploeckinger15} predict that compressive tidal forces can enhance the SF in TDGs compared
to classical dwarfs.
 Thus, given all these peculiarities,  TDGs are  unique  objects
 to investigate the universality of the SF law in  a  different environment than that of spirals or classical dwarf galaxies.

\begin{figure*}[h!]
\centering
\includegraphics[width=15cm]{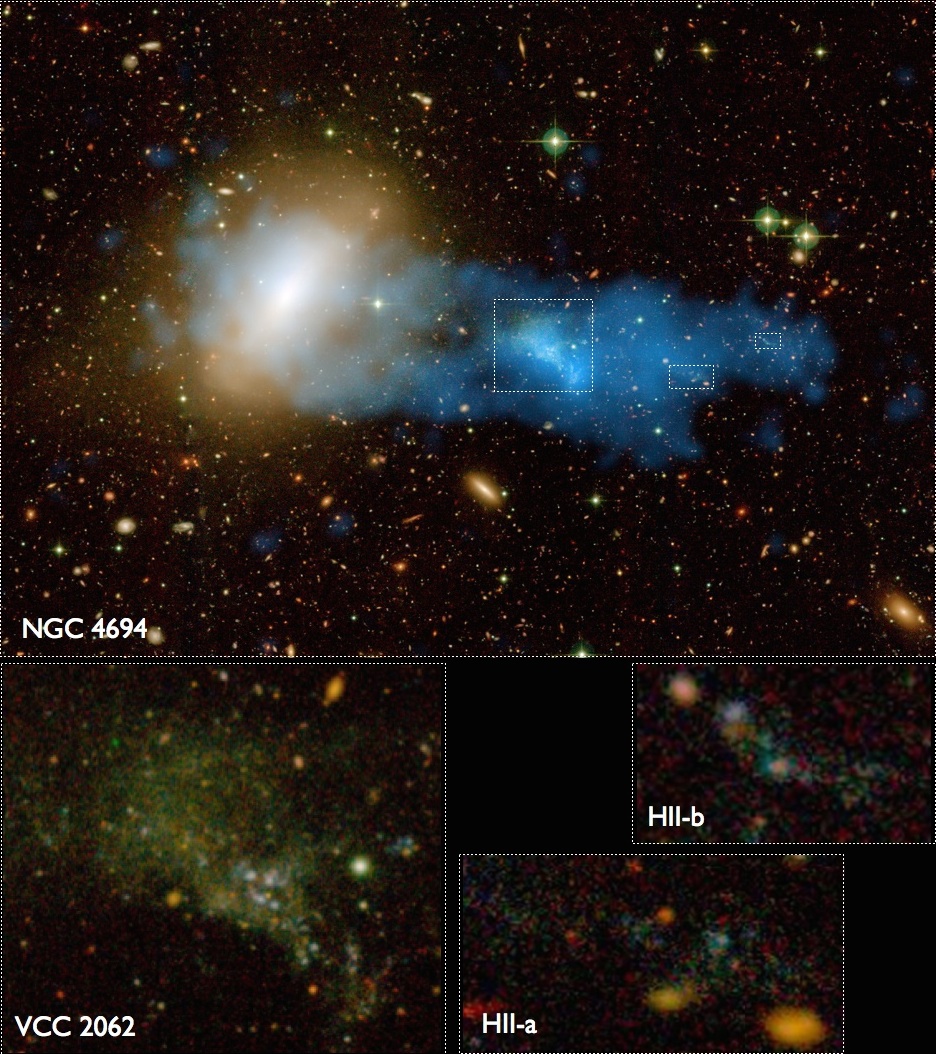}
%
%\includegraphics[width=15cm]{figures/NGC4694-VCC2062-HI.eps}
%\centerline{
%\includegraphics[width=7cm]{figures/TDG_optical_co_hi.eps}\quad
%\includegraphics[width=7cm]{figures/tail.eps}
 \caption{Optical false-colour image based on a combination of 
$u$, $g$ and $i$--bands from the NGVS of the system NGC4694 and VCC~2062. In blue  the HI
  emission is overlaid. At the bottom, enlarged images of 3 regions  within the HI tail are shown:  the TDG VCC~2062, and two 
  small SF regions at the end of the HI tail, HII-a and HII-b.
% {\it Lower left:} Blow-up of the region of \object{VCC~2062} {\bf drop HI and CO? Or drop only CO contours.
 % {\it Lower right:} Blow-up towards the end of the HI tip where a small SF region is visible.
% {\bf This figure as a whole still requires some improvement (labelling etc.}.
}
 \label{fig:ngc4694+vcc_optical+hi}
\end{figure*}

\object{VCC~2062} is a dwarf galaxy of extremely low surface brightness (central $\mu_{\rm v}$ = 25.5 mag/$\square$\arcsec) in the outskirts of the Virgo Cluster.
It is situated close to the perturbed, early-type galaxy NGC~4694 and is linked to it by a broad, long
HI plume (see Fig.~\ref{fig:ngc4694+vcc_optical+hi}).
It has been studied in detail  by \citet{duc07} with the conclusion 
that VCC~2062 is most likely an old  TDG and \object{NGC~4694} a merger remnant.
There are  two main observational
arguments that support this suggestion. Firstly,  the oxygen abundance (12+log(O/H) = 8.6, Duc et al.
2007) is about a factor of 10 above  the
value expected from its blue magnitude ($M_B = -13$~mag) and makes VCC~2062 a clear outlier
from the established magnitude-metallicity relation followed by classical dwarf galaxies \citep[e.g.][]{lee03}.
This discards the hypothesis that VCC~2062 is a preexisting dwarf galaxy % or used to be a large galaxy that has been totally disrupted 
and can be well  explained by the fact
that VCC~2062 is made of recycled gas from larger galaxies.
The second evidence is the low dynamical mass and dark-to-visible mass ratio which can be derived from the
atomic gas kinematics.
%The analysis of the HI data showed that VCC~2062 is the optical counterpart of
%a kinematically detached, rotating HI condensation that formed within the HI tail. A  second condensation
%exists at a distance of 2.6 kpc towards the south-west  \citep[called HI/SW in][]{duc07} with a slightly higher column density but
%very different kinematical properties. This part is not rotating and shows broader and more asymmetric 
%lines  and is most likely part of the underlying plume. Only weak optical emission
%was found in that region and was considered not to be part of  the TDG.
%Overall, It was concluded that VCC~2062 is most likely a TDG. 
Detailed modelling  of the HI distribution and kinematics and a comparison to the stellar and gaseous mass 
by Lelli et al. (2015) 
yielded a  dynamical-to-barionic mass ratio of 
$M_{\rm dyn}/M_{\rm bar} = 1.1 \pm 0.6$, i.e.  a value compatible with the absence of 
dark matter.
 The low dynamical mass discards
the hypothesis that VCC~2062 is the remnant of a pre-existing Low Surface Brightness (LSB) galaxy,
tidally disrupted by NGC~4694. Whereas gas and stellar material could have been lost
in such a disruption, a considerable decrease  of the non-barionic dynamical mass is much more difficult to envisage.

VCC~2062 is one of the closest known gas-rich TDG
%\footnote{
%The TDG candidate in the nearby M81 group, Holmberg 9, has no clear HI counterpart 
%({\bf Pierre-Alain, do you know a reference?})
%}
 and an ideal object for a detailed study of the relation between molecular gas
and SF in an extreme environment. 
In this paper we
present interferometric, high-resolution CO(1-0) data ($\sim 4$\arcsec, corresponding to a linear size of 330\,pc
at the distance of 17\,Mpc)  which we use to study the distribution
and the kinematics of the molecular gas. The main goal is to study
the SF law based on a comparison of the distribution of the molecular gas to the SFR measured
with different tracers. Our rich multiwavelength data-set furthermore allows us to model the
optical-to-infrared Spectral Energy Distribution (SED) to derive the present and past SFR and  the stellar mass.

\hspace{-4.cm}

%__________________________________________________________________

\section{The data}

This paper makes use of a wide set of data, partly form our own observations (CO(1-0)) and partly 
from the literature. The properties of the different data sets are summarized in Table~\ref{tab:overview_data}.

\begin{table}%[h!]
\caption{Summary of the observations used in this study.}
\label{tab:overview_data}
\begin{tabular}{lllll}
\noalign{\smallskip} \hline \noalign{\medskip}
Telescope & Instrument/ & $\lambda_0/\nu_0$ & FWHM & $\triangle_{\rm cal}$\tablefootmark{a}   \\
    & Filter & &  [$^{\prime\prime}$] &  [\%] \\
 \noalign{\smallskip} \hline \noalign{\medskip}
   GALEX &  FUX &    0.154~\mi &  4.2 & 4.7 \\
   GALEX &  NUV&    0.232~\mi  &  5.3 & 2.8 \\
   NGVS   &  $u$   & 0.381~\mi  & 1  & 4.7 \\
   NGVS   &  $g$   & 0.487~\mi  & 1  & 4.7\\
   NGVS   &  $i$   & 0.769~\mi  & 0.6  & 4.7 \\
   NGVS   &  $z$   & 0.887~\mi  & 1  & 9.6 \\
   \spitzer & IRAC & 3.5~\mi  & 1.6  & 10 \\
   \spitzer & IRAC & 4.5~\mi  & 1.7  & 10 \\
   \spitzer & IRAC & 5.8~\mi & 1.9  & 10 \\
   \spitzer & IRAC & 8.0~\mi  & 2.0  & 10 \\
   \spitzer & MIPS & 24~\mi  & 6  & 10 \\
   PdBI & $^{12}$CO(1-0) &  114.8 GHz& 4.3 $\times$ 4.1 \tablefootmark{b}   & - \\
    VLA & HI &  1.45~GHz & 14.7 $\times$ 14.3  &  -  \\
\noalign{\smallskip} 
\hline \noalign{\medskip}
\end{tabular}
\tablefoot{
\tablefoottext{a}{The calibration uncertainties are from \citet{morrissey07} (GALEX), \citet{raichoor11} (NGVS),  \citet{fazio04} (IRAC) and  
\citet{rieke04, engelbracht07,verley09} (MIPS).
We do not include calibration errors in the CO and HI data.}
\tablefoottext{b}{Beam FWHM of the taper90 datacuabe.}
}
\end{table}

% REference for FWHM: Fazio et al. 2004, Tab. 3

\subsection{CO(1-0) from  Plateau de Bure}

The observations were carried out between November 2010 and June 2011 in C and
D configuration with the Plateau de Bure Interferometer (PdBI) of the Institute de Radioastronomie Millimetrique (IRAM)\footnote{
IRAM is supported by INSU/CNRS (France), MPG (Germany) and IGN (Spain).
}.
We observed two overlapping positions with a central position of
RA 12:48:00.2, DEC 10:58:10 and offsets of (5\arcsec, 6\arcsec) and (-14\arcsec, 7\arcsec).
We observed the CO(1-0) line at a central sky-frequency of 114.842 GHz corresponding
to a recession velocity of 1120 \kms .
The November data suffered from cable phase inversion on antennas 5 and 6. The data were
 corrected {\em a posteri} for this problem.  
The observations were carried out under good weather conditions.
3C 273 was used as a phase and flux calibrator.
% {\bf what was used as amplitude calibrator?}.
%
%We used the xxxx receiver with the same setup in both polarizations ({\bf check}).
We used the correlator in two settings, one providing a velocity resolution of 0.82 \kms\ and a bandwidth
of 203 \kms ,
and the other at a velocity resolution of 6.2 \kms\ and a bandwidth of 2130 \kms. 
The data with the lower  resolution and larger bandwidth  were used to search for emission at 
high velocity offsets and as an attempt to detect the continuum emission. We did not find emission 
outside the velocity range explored with high-resolution data and neither did we detect 
any continuum emission. In the following we focus our analysis
on the high-velocity resolution data.

We reduced the data following  standard  procedures using the GILDAS\footnote{
http://www.iram.fr/IRAMFR/GILDAS
}
software. Since the emission in our maps was rather weak,
we used natural weighting to maximize  the sensitivity. 
The untapered map achieved an angular resolution of 3\farcs2 $\times$ 2\farcs4
but showed only weak emission. We therefore applied different tapers to the data
producing  maps at different angular resolutions
(4\farcs3 $\times$ 4\farcs1 for a taper to 90m and
7\farcs7$\times$ 6\farcs3 for a taper to 50m).
We will call these datacubes taper50 and taper90 in the following.
 We used the H\"ogbom cleaning procedure, and tried also the Clark procedure which gave no
 important differences. In the
cleaning process we identified and marked in each channel the areas  containing
emission. We used a loop gain of 0.1  (we also tried a loop gain of 0.2 without noticing 
important differences).  We cleaned down to 1.5\% of the peak emission (which corresponds to
about 0.5$\sigma$). We used the recommended truncation threshold of 
0.2 of the primary beam sensitivity. 
%We also tried a lower value of 0.1 in order to produce
%a slightly larger map which allowed us to investigate
%whether structures at the edge of the image
%were real.
We binned 2 channels of the high-velocity resolution data yielding a velocity resolution of 1.6 \kms. 
%For most of our
%analysis we  use the data cube  tapered to 50m because the signal-to-noise ratio is higher than
%for the 90m tapered cube, and the spatial resolution is still sufficient for a comparison
%to emission at other wavelengths.

%{\bf Something about moment maps, fitting of the spectra}

\subsection{Multiwavelength data from  \citet{duc07}}

We made use of the large collection of data presented in  \citet{duc07}, in particular
single-dish CO(1-0) and CO(2-1) data from the IRAM 30m telescope,
HI data, UV data in the FUV ($\lambda$ = 151 nm) and NUV ($\lambda$ = 227 nm) band from GALEX 
and an \halpha\ map. 
A detailed description of these observations and
their data reduction are presented in  \citet{duc07}. Here we only summarise
the most relevant aspects.

The HI data were retrieved from the VLA archive. 
They had originally been obtained as part of the VIVA project (project ID: AK563), a systematic survey of
HI rich objects in VIRGO \citep{chung09}.
The final, robustly weighted HI data cube that we use here  has a beam of 14\farcs7 $\times$ 14\farcs3
and  an rms noise of 0.6 mJy beam$^{-1}$, corresponding to
a detection threshold (assuming a 3 $\sigma$ detection across 3 channels) of
$9 \times 10^{19}$ cm$^{-2}$. The velocity resolution is 7.3 \kms.

The  CO spectra were observed with the IRAM 30m telescope. A map
with a spacing between pointings of 7\arcsec\ was obtained, covering roughly the extent of the 
HI cloud \citep[see Fig. 2 of][]{duc07}. The velocity resolution is 2.6 \kms\ for the CO(1-0)
spectra and 1.3 \kms\ for CO(2-1).

\subsection{Spitzer data}

We downloaded data obtained with  the \spitzer\ telescope \citep{werner04} from the
Spitzer Heritage Archive. The field containing VCC~2062 was observed with 
the Infrared Array Camera   \citep[IRAC,][]{fazio04}  and with the Multiband Imaging Photometer  \citep[MIPS,][]{rieke04}. 
The  images were processed with the Spitzer Super-Mosaic Pipeline and no further reduction was carried out.
VCC~2062 was detected in all IRAC bands and in the 24~\mi\ MIPS band. It was not detected by  MIPS at
70~\mi\ nor 160~\mi.

\subsection{Deep optical data}

The deep optical images used in this paper %and shown in Figs. 1 and 7 
were obtained at the Canada-France-Hawaii Telescope as part of the Next Generation Virgo Cluster Survey  \citep[NGVS,][]{ferrarese12}. NGVS used an observing  strategy and data reduction pipeline \citep[ELIXIR-LSB, Cuillandre et al., in prep.;][]{duc15} optimized for the detection of extended LSB objects. Images were obtained in the $u$, $g$, $i$, and $z$-bands.
 Thanks to the low limiting surface brightness achieved -- about  29 mag\,arcsec$^{-2}$ in the $g$-band --, a wealth of so far unknown faint 
 stellar structures  could be detected around NGC 4694 and give new insight in the history of this galaxy and its relation to VCC~2062.
 
 %  in particular a disturbed extended halo, and a tidal tail linking the early-type galaxy to VCC2062. They leave little doubt that the ETG has experienced in the past a major merger event. 

Fig.~\ref{fig:ngc4694+vcc_optical+hi} shows a combined three-colour image of both  objects in
which the SF regions in VCC~2062 are clearly visible. Both galaxies are connected by a long
and broad HI plume which extends to about twice the distance between them. Interestingly, towards
the end of this HI plume, two further, much smaller, SF regions can be seen.

\begin{figure}[h!]
 \centering
 \includegraphics[width=8cm,angle=0]{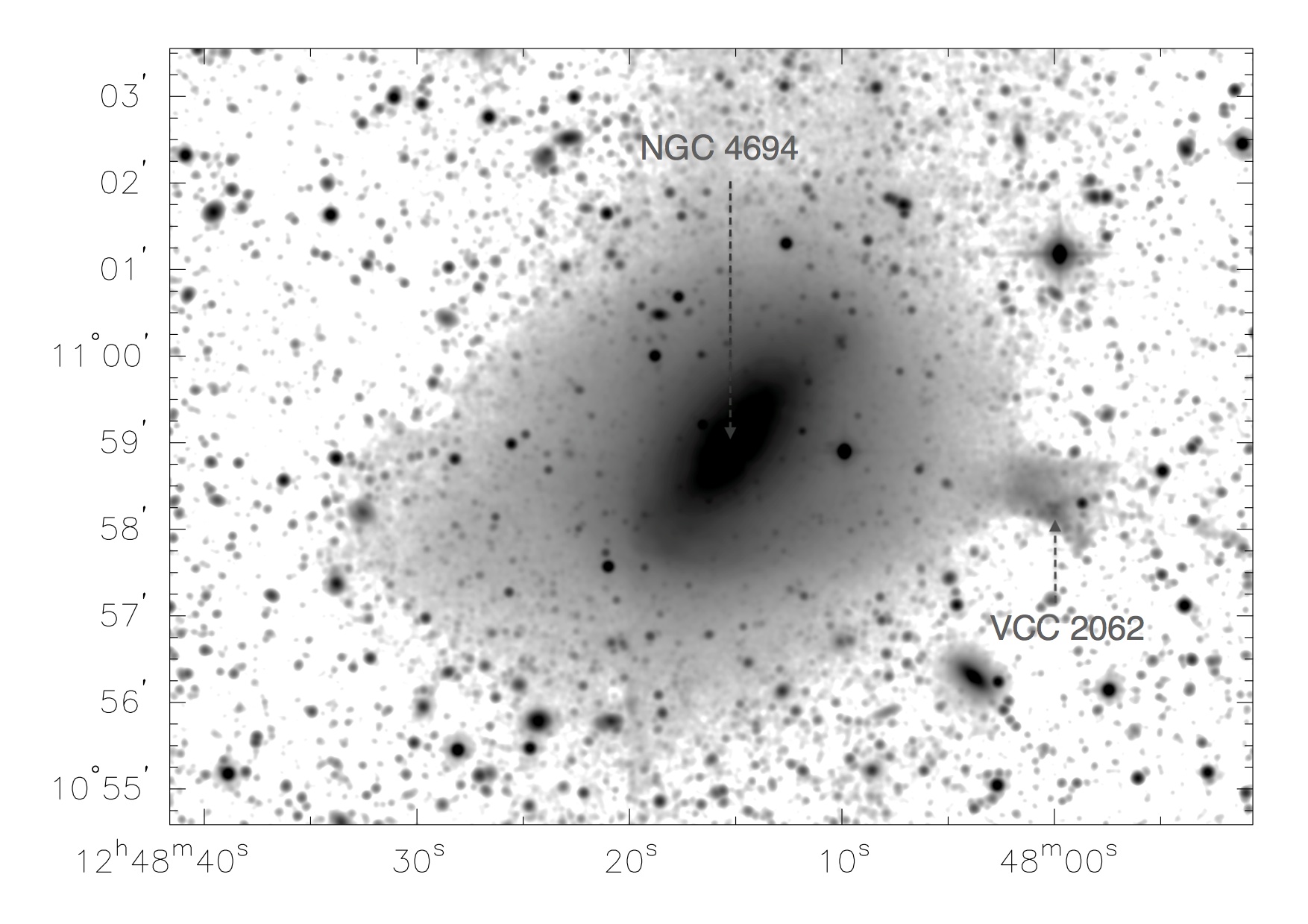}
 \caption{Deep $g$--band  image from the NGVS of NGC~4694 and VCC~2062. 
}
 \label{fig:ngc4694-deep-image}
\end{figure}

Fig.~\ref{fig:ngc4694-deep-image} shows an image in the $g$-band, in which the low brightness
structures are highlighted. The distorted and asymmetric appearance of NGC~4694 can be seen.
Very noticeable is the broad stellar bridge from this galaxy to VCC~2062.

%________________________________________ __________________________

\section{Properties of the molecular gas}

\subsection{Distribution and mass of the molecular gas}

\begin{figure}[h!]
 \centering
  \includegraphics[width=9cm]{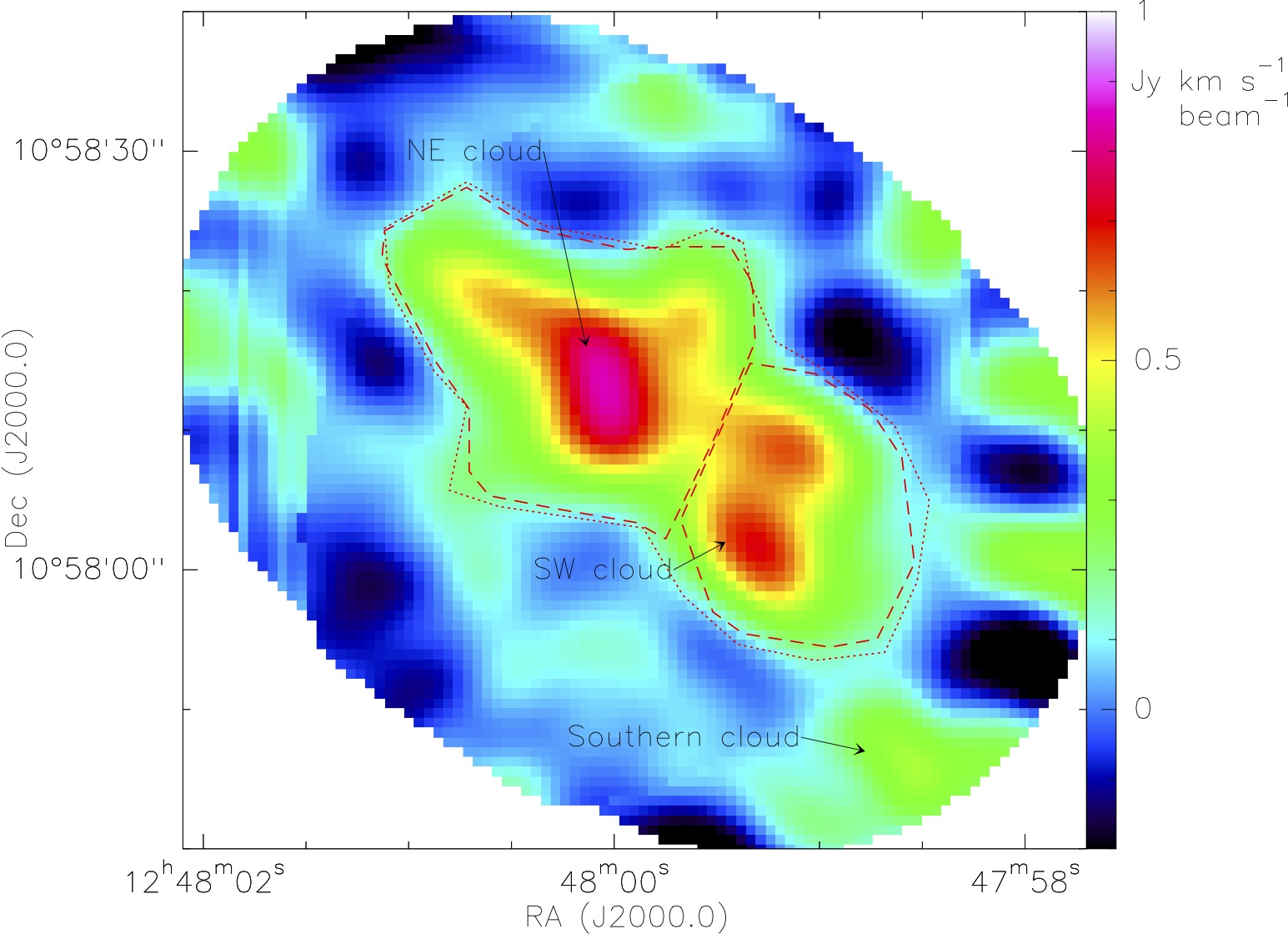}
\includegraphics[width=9cm]{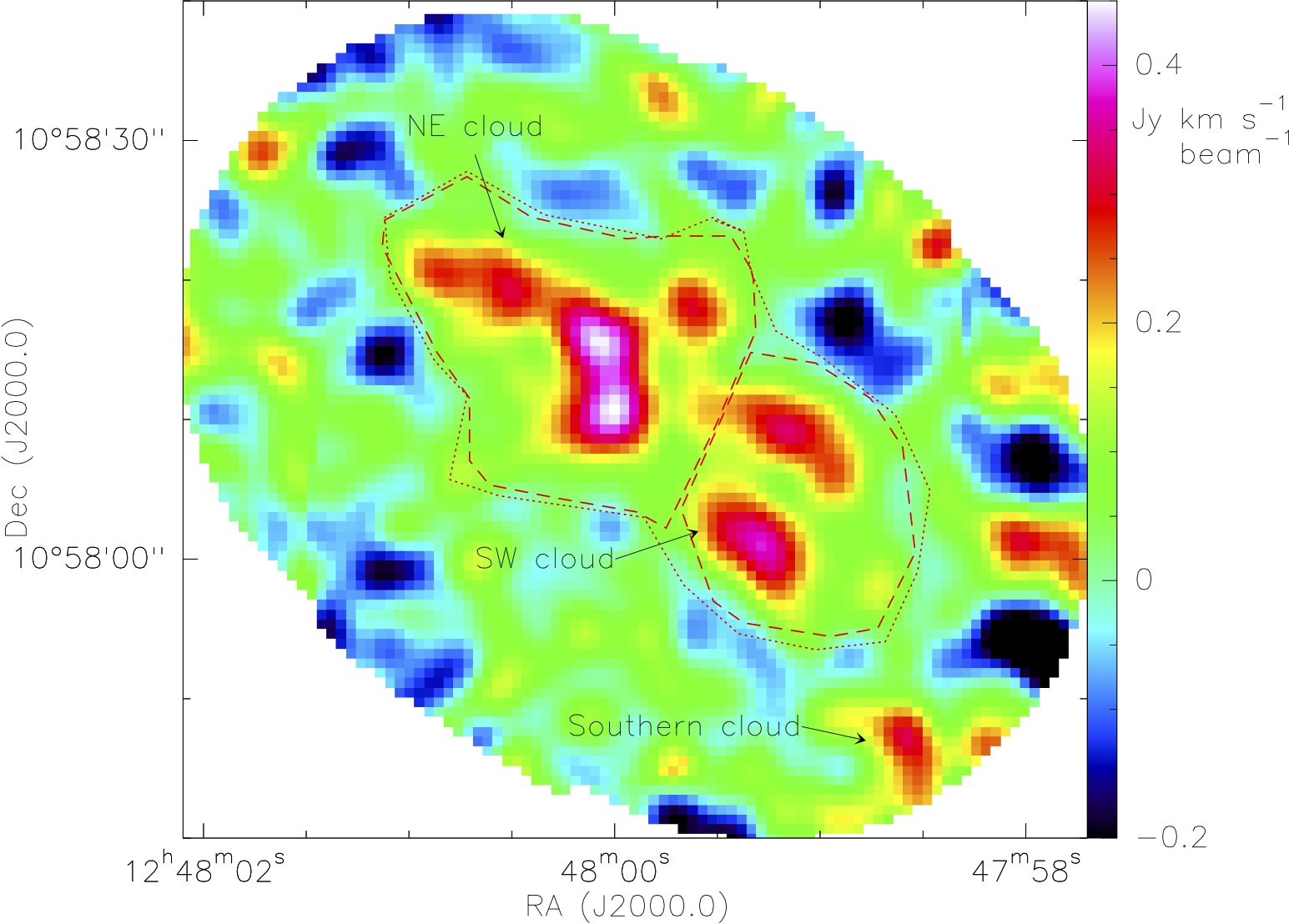}
 \caption{Velocity integrated (from 1129 to 1159 \kms) intensity of the CO(1-0) data cube.
  Different regions of emission are labelled.
 The contours indicate the apertures used to derive the  fluxes. 
  {\it Upper panel:} 
  data cube tapered at 50m (beam size 7\farcs7 $\times$ 6\farcs3),
 {\it Lower panel: }
data cube tapered at 90m (beam size 4\farcs3 $\times$ 4\farcs1).
  }
 \label{fig:co-mom0}
\end{figure}

Appendix A presents the channel maps of the taper50 data cube showing that 
emission is visible in the
velocity range between 1129 and 1159 \kms, distributed in 
various clouds and showing a velocity gradient along the NE-SW direction.
In Fig.~\ref{fig:co-mom0} we display the velocity integrated %(from 1129 to 1159 \kms) 
CO map.  
In the upper panel, two clearly separated regions of emission are visible which we call 
the NE and the SW clouds. In the very South, close to the edge of our imaged area, there
is a further region  (labelled ``Southern cloud"). We tested the significance of this
emission  by creating a larger map going out to 10\% of the primary beam sensitivity and inspecting
visually the spectra in this region. We concluded that the emission is real, but 
at the limit of significance. We will therefore not include this emission in our
quantitative analysis.
%
%We overlaid the contours of the velocity integrated emission from 1170 to 1185 \kms.
%This emission mostly coincided with the SW cloud. 
%We tested whether this emission could be due 
%to an artefact in only 1 channel by deriving the data cube at a higher (1.6 \kms ) 
%velocity resolution and confirmed the reality of this emission component in the SW region.
%
The higher-resolution image (lower panel),  shows  that  the NE,  and even more the 
SW, cloud consist of various smaller clouds. 

%\begin{figure}[h!]
% \centering
% \includegraphics[width=9cm]{figures/mean_emission_t90_paper.jpg}
% \caption{Velocity integrated intensity of the CO(1-0) data cube, tapered at 90m (beam size 4.28\arcsec $\times$ 4.12\arcsec ).
% The colour scale shows the result of the velocity integration from 1125 to 1160 \kms. 
% The labels indicate the naming of the different regions of emission.
% The contours indicate the aperture used for integration of the total flux.
%}
 %\label{fig:co-mom0-taper90}
%\end{figure} 

\begin{figure}
 \centering
 \includegraphics[width=7.5cm]{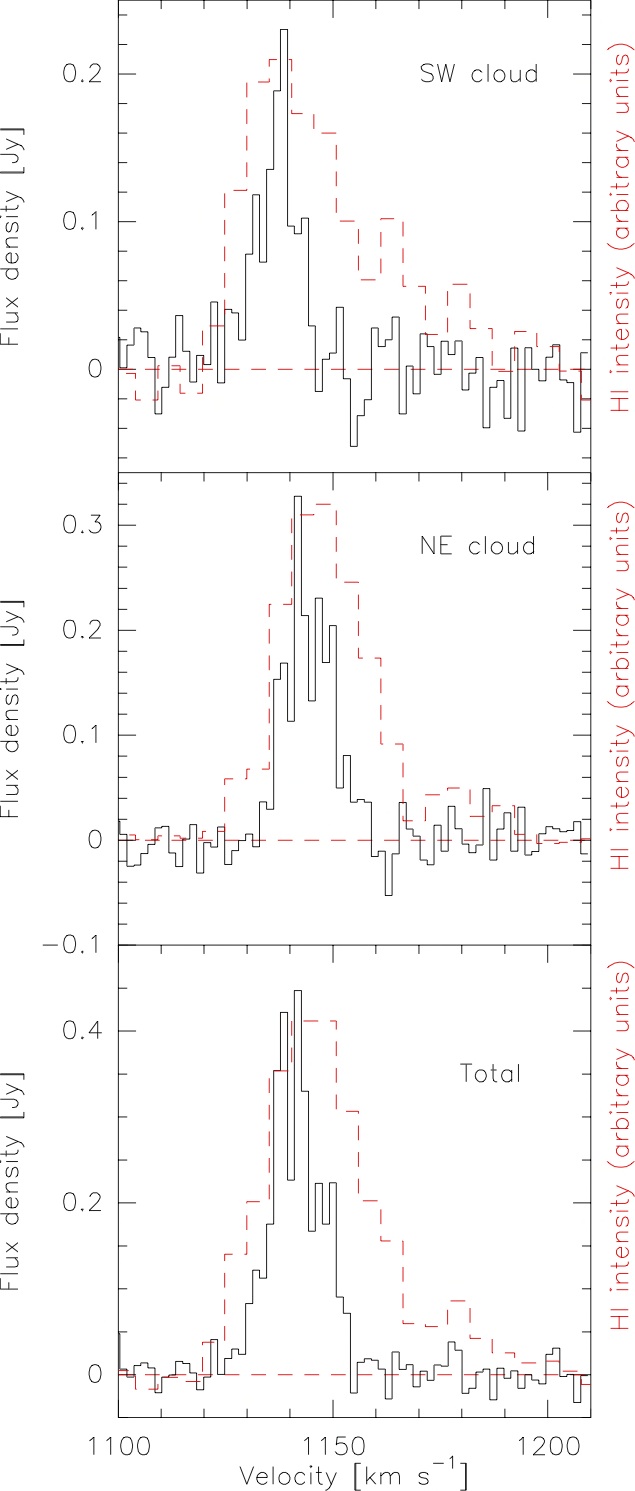}
\caption{CO spectra (black lines), integrated over different regions (see Fig.~\ref{fig:co-mom0} for the
corresponding apertures), overlaid with the HI spectra (red dashed lines) in the same areas.
 }
 \label{fig:spectra}
\end{figure}

Based on this distribution of the CO, we defined different regions in the galaxy, corresponding
to the NE and SW cloud and the total emission (see Fig.~\ref{fig:co-mom0}).
For each region, we determined the integrated spectrum which is shown in Fig.~\ref{fig:spectra}.
The line widths in the two  regions are $\sim$ 20 \kms (SW) and  $\sim$~25  \kms\ (NE).
We then determined the velocity integrated flux of the spectra by integrating over  channels 
with emission. The error was calculated as
$\triangle S = \sigma \sqrt{\triangle V dV} $ where $dV$ is the width of a channel (1.6 \kms),  
$\triangle V$ is the velocity range over which the integration was carried out %(35~\kms) 
and $\sigma$ is 
the rms noise level, determined  in the part of the spectrum where no emission was detected.  
%
%derived from the different regions  is listed in
%Table~\ref{tab:flux_mass_regions}. 
The values are measured from the taper90 map, but
the agreement with the taper50 map is within $\sim$ 10\%.
The results are listed in Table~\ref{tab:flux_mass_regions}.
 We can compare the value to the total flux obtained over the same area with  the IRAM 30~m telescope of 
7.5 Jy \kms  \citep{duc07}, with an estimated error of about 20\%. Thus, the interferometric observations 
 yield   about (70$\pm$15)\%  of the flux from
 single-dish observations, showing that at most a small fraction of smoothly distributed,  diffuse
CO may be resolved out.

We calculated the mass of the molecular gas from the PdBI CO fluxes as
\begin{equation}
M_{{\rm mol}} = 1.05\times 10^4 D_{\rm Mpc}^2 S_{\rm CO(1-0)}\, M_\odot,
\end{equation}
where $D_{\rm Mpc}^2$ is the distance in Mpc and $S_{\rm CO(1-0)}$ is the CO(1-0) flux
in Jy \kms 
\citep[see][]{bolatto13}.
This expression is based on a conversion factor of \xco\ = $2\times 10^{20}$ cm$^{-2}/$(K\,\kms) and
includes a mass fraction of helium of 1.36. %, and assumes that the redshift $z\approx 0$.
The  molecular gas masses are listed in Table~\ref{tab:flux_mass_regions}, together with 
the molecular mass surface densities, which were calculated by dividing the mass by the area over which
the CO flux was integrated.

\begin{table*}%[h!]
\caption{CO(1-0) velocity  integrated flux (from 1129 to 1159 \kms), S$_{\rm CO}$, molecular gas mass, 
mean molecular surface density, atomic gas mass, mean atomic surface density, and molecular gas mass fraction
in different regions.}
\label{tab:flux_mass_regions}
\begin{tabular}{llllllll}
\noalign{\smallskip} \hline \noalign{\medskip}
Region & Area & S$_{\rm CO}$ & \mmol & $\Sigma_{\rm mol}$ & \matom\tablefootmark{b}  & \sigmaatom & \mmol/\mgas\\
   & [kpc$^2$] & [Jy \kms]  & [10$^6$ \msun]  & [\msun pc$^{-2}$]  & [10$^6$ \msun]  & [\msun pc$^{-2}$]  & \\
\noalign{\smallskip} \hline \noalign{\medskip}
% with heium 
   NE &  2.9  &    3.4$\pm$0.2  &  10.3$\pm$ 0.6&   3.6$\pm$0.2  & 26$\pm$ 3&  8.9$\pm$0.9 & 0.28$\pm$0.03\\
   SW &   1.7 &     2.0$\pm$0.2 &  6.1$\pm$0.6 &  3.5$\pm$0.4 & 13$\pm$1 &  7.9$\pm$0.8 & 0.32$\pm$0.05 \\
     total\tablefootmark{a} &  5.2  &   5.4$\pm$0.2  & 16.4$\pm$0.6  &  3.2$\pm$0.1 & 46$\pm$5  & 8.9$\pm$0.9 & 0.26$\pm$0.03\\
% no heiums
%    NE &  2.9  &    3.8  &  8.6 &   3.0 \\
 %   SW &   1.7 &     2.3 &  5.2 &  3.0 \\
%  both  &  5.2  &   6.2  & 14.1  &  2.7\\
\noalign{\smallskip} 
\hline \noalign{\medskip}
\end{tabular}
\tablefoot{
\tablefoottext{a}{"Total" refers to the CO-emitting area for both the molecular and the atomic gas mass, not to the total HI area of the TDG (see Fig.~\ref{fig:pv-hi-co-lelli}).}
\tablefoottext{b}{For consistency with the molecular gas, we include in the atomic gas mass a helium fraction of 1.36.}
}
\end{table*}

% The error in MHI was estimated from a comparison of the results from robust and natural
% weighting in  order to estimate uncertainty by large spatial resolution.

\subsection{Comparison of CO and HI emission}

\begin{figure*}[H]
 \centering
 \includegraphics[width=18cm]{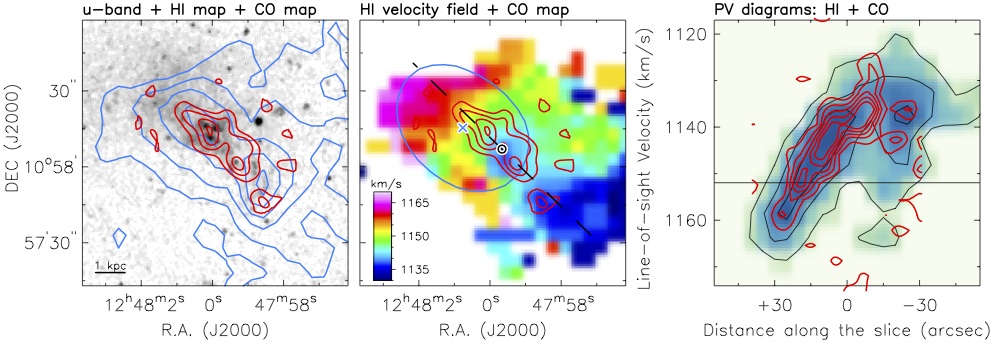}
 \caption{\textit{Left panel}: $u$-band image overlayed with  the CO (red contours) and HI emission (blue contours). 
 CO contours are at $\sim$0.2, 0.4, 0.6 and 0.8 Jy \kms\ beam$^{-1}$, corresponding to \sigmamol $\sim$ 2. 4, 6 and 8 \msun\ pc$^{-2}$. 
 % tansformation: Mmol = S * 1.37*7.83e3*dist**2
 %           Area = 1.13*beam1*beam2*0.29*0.29*dist **2  (beams in arcmin, area in kpc)
 % with S = 0.2, beam 1= 7.6|60, beam 2= 7.0|60  >-->  Sigma_mol = 0.2*1.37*7.83e3*3600*1.e-6|(7.0*7.6*0.29*0.29)
HI contours are at \sigmaatom\ $\sim$ 2, 4, 6, and 8 \msun\ pc$^{-2}$. %1.5, 3.0, 4.5, and 6 \msun\ pc$^{-2}$. 
 The bar to the bottom-left corner corresponds to 1 kpc. \textit{Middle panel}: HI velocity field overlayed with the CO emission (contours are the same as in the left panel). The blue cross and 
ellipse  illustrate, respectively, the centre and extent of the HI disc investigated by \citet{lelli15}. The black circle and dashed line show, respectively, the centre and orientation of the slice used to obtain the  PV diagrams. \textit{Right panel}: PV diagrams obtained from the HI and CO cubes along the dashed line in the middle panel. 
 %CO contours (in black) are at 10, 20, 30, and 40 mJy beam$^{-1}$. HI contours (in blue) are at 1.2, 2.4, and 3.6 mJy beam$^{-1}$. 
Contours range from 3 to 15 $\sigma$ in steps of 3$\sigma$, where $\sigma = 4.5$ mJy beam$^{-1}$ for CO data (red) and $\sigma = 0.6$ mJy beam$^{-1}$ for HI data (black).The horizontal line corresponds to the systemic velocity of the HI disc.
 }
 \label{fig:pv-hi-co-lelli}
\end{figure*}

Figure~\ref{fig:pv-hi-co-lelli}  compares the distribution and kinematics of CO and HI emission. The left panel shows an $u$--band image overlaid with both the total CO map at $\sim$7$''$ resolution (red) and the HI map at $\sim$14$''$ resolution (blue). The CO cloud to the NE coincides with a strong HI peak (\sigmaatom $\sim$8 M$_{\odot}$ pc$^{-2}$)
and several stellar clusters. The CO cloud to the SW, instead, lies between two HI peaks; faint star clusters are present in this region.
The molecular gas fraction is around 30\% everywhere (see Table~\ref{tab:flux_mass_regions}) which is in the range of what \citet{casoli98} found for  spiral galaxies of type Sa-Sc, and much higher than
their result for Sd spirals and irregulars (less than 5\%).

The middle panel shows the HI velocity field \citep[from][]{lelli15} overlaid with the total CO map. \citet{lelli15} analysed high-resolution HI data of VCC 2062 and found that the HI emission can be described by a rotating disc model. The blue cross and ellipse in Fig.~\ref{fig:pv-hi-co-lelli} illustrate, respectively, the centre and extension of the HI disc as derived by \citet{lelli15}. Clearly, the CO emission is displaced to the SW of the kinematic centre, lying on the approaching, blue-shifted side of the HI disc.
The NE half of the HI disk contains only older stars.

The right panel compares Position-Velocity (PV) diagrams derived from the HI and CO cubes (red and black contours, respectively). 
%The slice centre and orientation are indicated, respectively, by the black circle and dashed line in the middle panel of Fig.~\ref{fig:pv-hi-co-lelli}. 
There is excellent agreement between HI and CO kinematics. The HI gas has a broader line width
than the CO gas (see also Fig.~\ref{fig:spectra}    for a direct comparison of the CO and HI spectra),
which is most likely due to the 
%seems to have higher velocity dispersion than the CO gas, although this may be an observational effect due to the
 lower velocity and angular resolutions of the HI data. The horizontal line indicates the HI systemic velocity: clearly the CO mainly probes  the kinematics on the approaching side of the disc. The shape of the CO PV diagrams suggests a rising rotation curve that perhaps flattens in the outer parts, as it is often seen in dwarf galaxies \citep[e.g.][]{swaters09,lelli12a,lelli12b}. We note, however, that these PV diagrams do not correspond exactly to the disc major axis (they were chosen to go through the peaks of the CO emission), hence they do not capture the full disc rotation  \citep[cf.  Fig. 5 of][]{lelli15}. 
Towards the very SW, outside the ellipse showing the extension of the HI disk, the HI emission is not consistent anymore with a rotating disc; it is representative of the underlying tidal debris, showing broad and asymmetric HI line profiles.

%We conclude that the SW approaching side of the HI disc contains CO emission and young SF regions, which appear to be rotating around the HI kinematic center. The NE receding
% side, instead, contains only HI emission and older stars. Further towards the SW direction, the HI emission is not consistent anymore with a rotating disc; it is representative of the 
%underlying tidal debris, showing broad and asymmetric HI line profiles.

%%%%%%%%%%%%%%%%%%%%%%%%%%%%%%%%%%
\subsection{Comparison of CO with the distribution of SF regions}

\begin{figure}[h!]
 \centering
 \includegraphics[width=8cm]{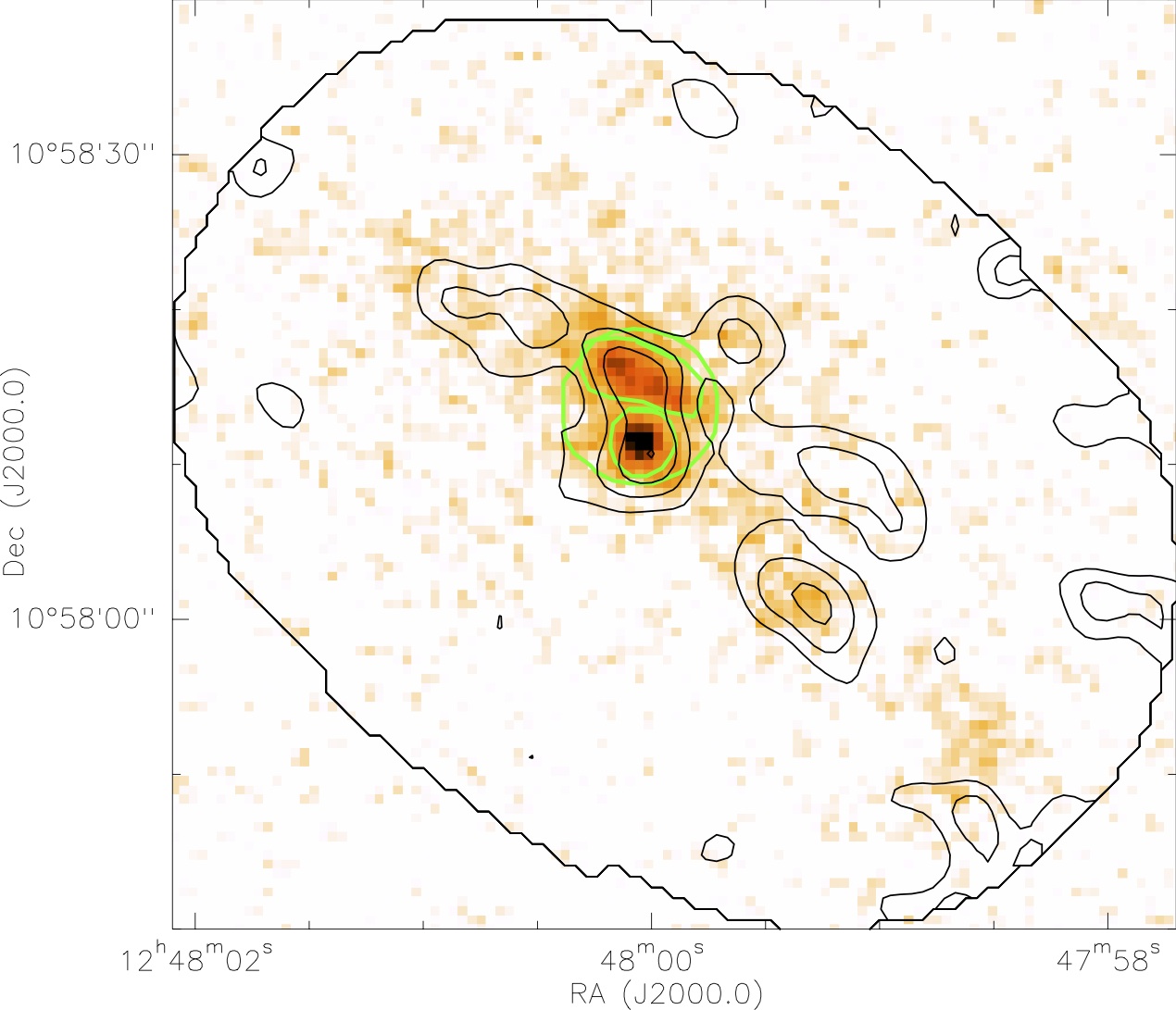}
 \caption{Overlay of IRAC 8~\mi\  data (colour) and CO contours (black contours), starting at 0.15 Jy \kms\ beam$^{-1}$ ($\sim 3\sigma$)
 and incrementing in steps of  0.1 Jy \kms\ beam$^{-1}$.  The green contours show the apertures used
 for the measurement of the SFR and molecular gas. The outer black line shows the Field of View of the PdBI observations.}
 \label{fig:irac-co}
\end{figure} 

In Fig.~\ref{fig:irac-co}  we show an overlay
of the CO emission (contours) with  the  8~\mi\ emission.
%and in Fig. \ref{fig:mips-uv-co}  with 24~\mi\ and UV. 
The NE CO cloud coincides with considerable SF, visible in all tracers, whereas at the position of
the SE cloud weaker 8 \mi. % and very little \halpha\ emission is present. 
SF traced by 8 \mi\ is also present close to the emission of the tentative weak southern CO cloud,
close to the edge of our  field of view.

%\begin{figure}[h!]
% \centering
% \includegraphics[width=8cm]{figures/overlay_24mu_uv_co_paper.jpg}
% \caption{Overlay of MIPS 24~\mi\  data (colour), UV\ (green contours) and CO contours (black contours), starting at 0.15 Jy \kms\ beam$^{-1}$ ($\sim 3\sigma$)
% and incrementing by 0.15 Jy \kms\ beam$^{-1}$.}
% \label{fig:mips-uv-co}
%\end{figure} 

%%%%%%%%%%%%%%%%%%%%%%%%%%%%%%%%%%
\section{Spectral Energy Distribution and modelling}

\subsection{Photometry}
\label{sec:photometry}

%\begin{figure}[h!]
% \centering
% \includegraphics[width=8cm]{figures/ngvs_u.eps}
% \caption{Apertures used for the photometry overlaid on a deep u-band  image from the NGVS of \object{VCC~2062}. 
% The two black apertures  show the regions used for the CIGALE  modelling (entire galaxy (full line), and
% a quiescient area with no young SF (dotted line)). The dashed  blue contour gives, for reference, the region in which CO was detected
% by the PdBI observations.}
% \label{fig:uband-contours}
%\end{figure} 

We made use of the multiwavelength maps available for VCC 2062 
and determined the Spectral Energy Distribution (SED)  from the FUV to 
24~\mi . We used  the same aperture 
%(see Fig.~\ref{fig:uband-contours}, 
for all bands and carried out the photometry in 2 areas, (i) an aperture
encompassing the entire object visible in the optical, including an extended area 
 with weaker emission and no signs of substantial young SF, as traced for example by  \halpha\  or 8~\mi\ emission (see the lower left image
 in Fig.~\ref{fig:ngc4694+vcc_optical+hi}), and (ii) an aperture encompassing the total CO emission 
%Apart from this global aperture, we carried out the photometry only for the
%north-eastern part in order to probe the stellar populations in a region without strong,
%recent SF 
(the same aperture as in Fig.~\ref{fig:co-mom0}).
 For each image,
we subtracted the average sky  background measured in close-by regions  outside of
the main aperture.
For the Spitzer images we applied, following the IRAC and Spitzer handbooks
 aperture correction factors of 0.95, 0.97, 0.88 and 0.83 for the IRAC bands 1 to 4 and 
a factor of 1.12 for the MIPS 24 \mi\ band.
For the NGVS data no aperture corrections were necessary due to their high angular resolution
compared to the aperture size. 
For the GALEX images we estimated the aperture correction from
the 8~\mi\ image which shows a similar structure. The aperture
correction was derived as the ratio between the flux of  the original 8~\mi\ map to the
flux from the 8~\mi\ map convolved to the GALEX resolution. We derived values for the
aperture correction of 2\%  for the FUV and 3\% for the NUV emission.

We determined the error as the quadratic sum of 3 components:
(i) The photometric error due to the noise in the image. This component was
usually negligible. (ii) The error due to variations
in the background level, estimated from the differences of the mean values in the various background
apertures. (iii) Calibration uncertainty as listed in Table~\ref{tab:overview_data}.

%which was adopted to be
%4.7~\% (0.05 mag, FUV) and 2.8~\% (0.03 mag, NUV) \citep{morrissey07}, 4.7~\% (0.05 mag) for the NGVS $g$, $i$ and $z$--band and 
%9.6~\% (0.1 mag) for the NGVZ $u$--band \citep{raichoor11},
%10 \% for the IRAC bands \citep{fazio04} and for MIPS 24 \mi\  \citep{rieke04,engelbracht07,verley09}.

The fluxes derived are  listed in Table~\ref{tab:sed}. They are corrected for a Galactic extinction
of  $A_V = 0.11$,  
%obtained from NED, 
based on the \citet{schlafly11} recalibration of the \citet{schlegel98}
extinction values and extrapolated using the  \citet{cardelli89} extinction law with $R_V = 3.1$.

\begin{table}%[h!]a
\caption{Fluxes for the total emission in various bands}
\label{tab:sed}
\begin{tabular}{llll}
\noalign{\smallskip} \hline \noalign{\medskip}
Band & $\lambda$ & \multicolumn{2}{c}{Flux} \\
  & [\mi ] & \multicolumn{2}{c}{[mJy]}  \\
\noalign{\smallskip} \hline \noalign{\medskip}
 &  & total galaxy  & CO area \\
\noalign{\smallskip} \hline \noalign{\medskip}
FUV (GALEX)  & 0.154  & 0.055  $\pm$ 0.005 & 0.039  $\pm$ 0.004\\
NUV (GALEX)  & 0.232  &0.083  $\pm$ 0.005 &0.051  $\pm$ 0.003 \\
u (NGVS)  & 0.381  &0.19 $\pm$0.02 &0.10 $\pm$0.01\\
g (NGVS)  &  0.487 & 0.32$\pm$ 0.03 & 0.15$\pm$ 0.01 \\
i (NGVS)  &  0.769  & 0.35 $\pm$ 0.03 & 0.17 $\pm$ 0.01\\
z (NGVS)  & 0.887  &0.33 $\pm$ 0.06 &0.15 $\pm$ 0.02\\
IRAC1 (\spitzer)  &3.5  &0.23 $\pm$ 0.03 &0.14 $\pm$ 0.02\\
IRAC2 (\spitzer)  &4.5 & 0.19 $\pm$0.02 & 0.10 $\pm$0.01\\
IRAC3 (\spitzer)  &5.8 &  0.83  $\pm$ 0.35 &  0.59  $\pm$ 0.14\\
IRAC4 (\spitzer) & 8.0  &1.9  $\pm$ 0.4 &1.3 $\pm$ 0.2\\
MIPS1 (\spitzer) &24  & 1.3  $\pm$ 0.7 & 1.0  $\pm$ 0.3\\
\noalign{\smallskip} 
\hline \noalign{\medskip}
\end{tabular}
\end{table}

\subsection{SED modelling with CIGALE}

To estimate physical properties of VCC~2062, we modelled the SED with 
the latest release of the CIGALE code\footnote{http://cigale.lam.fr}
%\footnote{\url{http://cigale.lam.fr}} 
(version number 0.8.1, Boquien et al. in prep.).
We built a grid of models,
 including different SF histories, stellar populations, attenuations (shape of the attenuation curve
 and normalisation), and emission by dust. 
 The full grid is made of 20,966,400 models.
In more detail, these components cover the
 following parameter space:
\begin{enumerate}

\item Guided by the high UV/\halpha\ luminosity ratio, we modelled the SF history by a double decaying exponential. The first 
exponential represents the long term SF of the object, starting 
13~Gyr ago and with an $e$--folding time ranging from 1~Gyr to 8~Gyr in steps of 1~Gyr. The second 
exponential represents the latest star--forming episode having started 
between 50~Myr and 500~Myr ago, with an $e$--folding time from 50~Myr to 
500~Myr, both in steps of 50~Myr. This second episode is assumed to have formed between 0.1\% and 50\%
(the sampling being 0.1\%, 0.5\%, 1\%, 5 \%, 10\%, and afterwards increasing in steps of 5\% until 50\%)
of the total stellar mass.

\item The stellar populations are assumed to have a constant, solar metallicity of
Z=0.02. This metallicity is appropiate for VCC~2062 which has
12+log(O/H) = 8.6 \citep{duc07}, close to the Solar metallicity of 12+log(O/H) = 8.66 \citep{asplund05}.

\item We adopt a    \citet{chabrier03} Initial Mass Function (IMF), which is very similar to the Kroupa IMF.

\item The stellar populations are attenuated with a power--law--modified 
starburst extinction law ($A(\lambda)=A(\lambda)_{SB}\times(\lambda/550~{\rm nm})^\delta$, with 
$-0.5\le\delta\le0.0$ (in steps of 0.1)). The $E(B-V)$ reddening of stars younger than 10~Myr is 
comprised between 0.1 mag and 0.75 mag (increasing in steps of 0.05 mag). 
Young stars that have not yet broken out completely from their birth clouds tend to be more attenuated 
than the old stellar population \citep[e.g.][]{calzetti13}. In order to take this into account, we apply a 
reduction factor (0.25, 0.50, or 0.75) to the reddening of old stars. 

\item The emission of dust is computed from the absorbed energy from the 
UV to the near--infrared through an energy balance principle, modelled using 
the \citet{dale14} template, with the $\alpha$ parameter varying between 
0.5 and 4.0, in steps of 0.5.
\end{enumerate}

Finally, the physical properties such as the SFR, the stellar mass, or the 
attenuation are estimated from the full grid of models through a Bayesian--like 
analysis of their respective marginalised probability distribution function.

\begin{figure}[h!]
 \centering
 \includegraphics[width=7cm,angle=90]{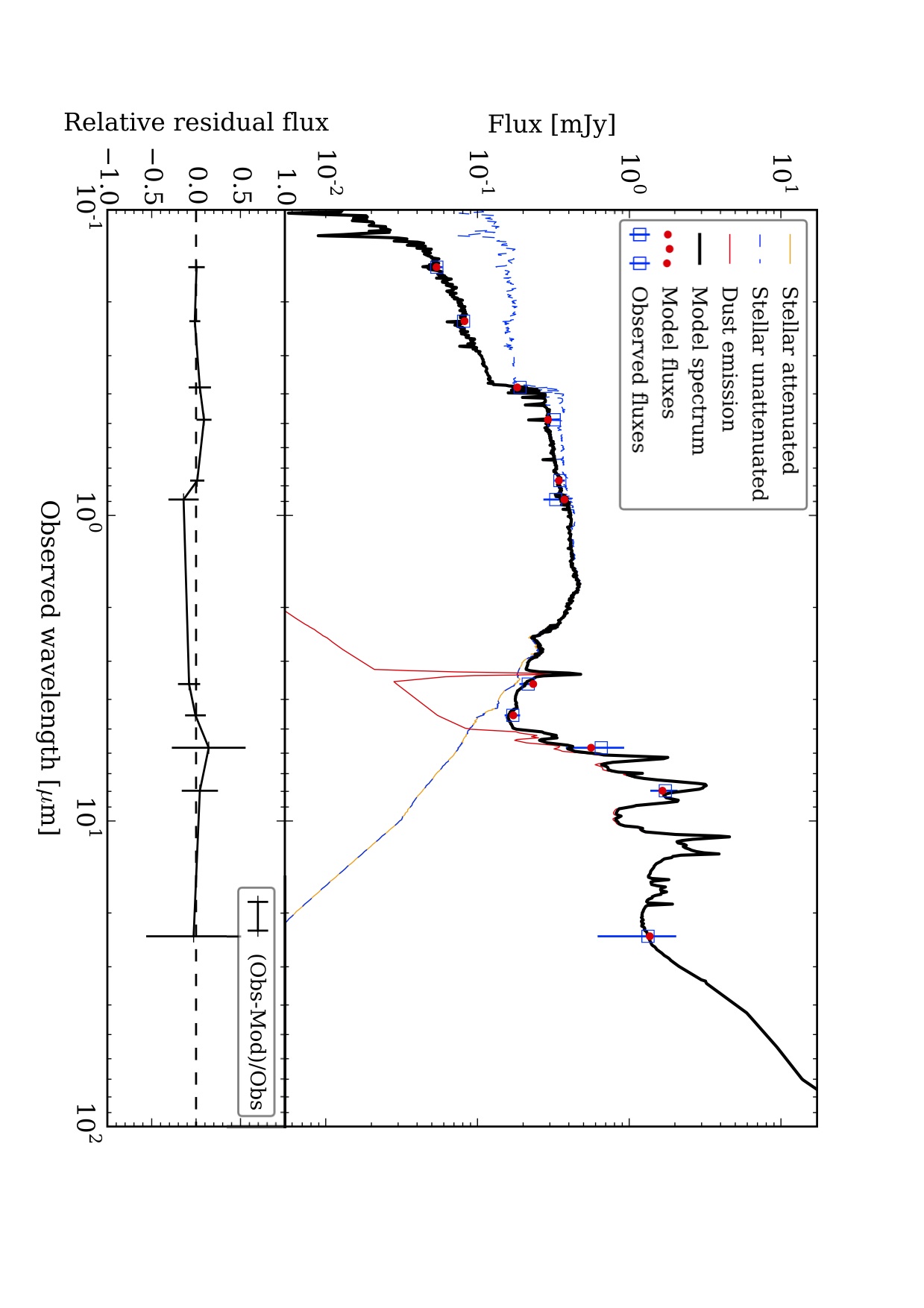}
 \caption{Observed SED of VCC~2062,  together with the best fit derived with CIGALE. 
%The error bars refer to 1$\sigma$. 
In the bottom panels the residuals from the comparsion between model and observations
are shown.
 }
 \label{fig:cigale-fit}
\end{figure}

We show the best fit for the total galaxy in Fig.~\ref{fig:cigale-fit}. The most relevant best-fit parameters  are listed in
Table~\ref{tab:output_best_fit} (both for the total galaxy and for the CO-emitting region).
 The stellar emission was fitted by a  combination of a burst and an underlying
older component. The mass fraction in the burst ($\sim$ 40\%) is found to be considerable but the exact
value is not well constrained. 
The age of the burst of about 0.3 Gyr is consistent with the expected age of the merger (several hunddred Myr). 
The SFR has been declining by a factor of about 1.4 during the last 100 Myr.

The SED of VCC~2062 has been modelled before by \citet{boquien10}.
 The present analysis includes two major improvements. (i) The quality of the data set 
 has improved substantially. Apart from a higher image quality in the optical range, the data set now includes 
 high-quality data in the $i$ and $z$  bands which were absent or very poor in \citet{boquien10}.
 (ii) The modelling is done 
 with the CIGALE code instead of PEGASE which was used in \citet{boquien10}. The most
 relevant improvement in CIGALE is that the fitting process relies on a Bayesian
 estimate of the marginalised probability distribution functions, allowing more reliable estimates of 
 the physical properties and the associated uncertainties. Furthermore, the entire SED from the UV
 to the IR is modelled consistently now, with the energy absorbed in the UV being re-emitted in the
 IR by the dust. This allows us to make use of the mid-IR emission
which is crucial to break the age-attenuation degeneracy of the object. 
The results obtained here remain broadly consistent with those  from \cite{boquien10},
but represent a considerable improvement, for example  with respect to the the SFR  which could not be determined
reliably in the previous study.

\begin{table}%[h!]
\caption{Best-fit parameters derived with the CIGALE SED-fitting}
\label{tab:output_best_fit}
\begin{tabular}{lll}
\noalign{\smallskip} \hline \noalign{\medskip}
Parameter& total galaxy & CO area \\
\noalign{\smallskip} \hline \noalign{\medskip}
present SFR      ($10^{-3}$  \msun\  yr$^{-1}$)   & 3.9 $\pm$  0.9  & 3.0 $\pm$  0.4 \\
% North:  3.57e-4 msun/yr
past SFR\tablefootmark{a} ($10^{-3}$ \msun\ yr$^{-1}$)   & 5.4 $\pm$  1.0  & 3.7 $\pm$  0.7 \\
% North:  67.26-4 msun/yr
total stellar mass  ($10^{6}$ \msun) & 7.1 $\pm$  2.0   & 3.1 $\pm$  0.9   \\
%mass fraction in the burst & 9\% \\
age of the burst  (Myr)  & 376 $\pm$  99 & 313 $\pm$  86  \\
$A_V$    (mag)  &   0.24 $\pm$ 0.06  &   0.28 $\pm$ 0.06 \\
$A_{FUV}$  (mag)  &  1.25 $\pm$ 0.17  &  1.31 $\pm$ 0.15\\
\noalign{\smallskip} 
\hline \noalign{\medskip}
\end{tabular}
\tablefoot{
\tablefoottext{a}{SFR averaged over the past 10$^8$ yr.}
}
\end{table}

% OLD VALUES
%\noalign{\smallskip} \hline \noalign{\medskip}
%present SFR                                    & (2.43 $\pm$  0.68) $10^{-3}$  \msun yr$^{-1}$ \\
% North:  3.1e-4 msun/yr
%averaged SFR (over past $10^8$ yr)  & (3.91 $\pm$  0.97) $10^{-3}$  \msun yr$^{-1}$ \\
% North:  6.8e-4 msun/yr
%total stellar mass & (1.27 $\pm$  0.21) $10^{7}$  \msun  \\
%mass fraction in the burst & 9\% \\
%age of the burst   & 293 $\pm$  92 Myr\\
%$A_V$      &   0.15 $\pm$ 0.04 mag \\
%$A_{FUV}$  &  0.86 $\pm$ 0.16 mag \\
% END

%%%%%%%%%%%%%%%%%%%%%%%%%%%%%%%

\section{The SF  law in VCC 2062}
\label{sect:ks-law}

\subsection{Calculation of the star formation rate}

The high-resolution CO data allows us to study the SF law separately
for the NE cloud and SW cloud, as well as for the region encompassing the entire molecular gas
emission. Hence, we are probing SF on a $\sim$~kpc scale (the linear sizes of the NE and SW regions are 2.5 and
1.6~kpc, respectively).
We calculated the SFR using different tracers which 
 allows us to obtain a  complete picture of the SF history.
Whereas \halpha\ is emitted from the most massive, ionizing stars and traces the very recent
SFR ($\sim$ 10\,Myr),  the UV radiation is emitted mostly by  stars of several solar masses and probes the
SFR  over the past $100 - 200$\,Myr.  Both emissions need to be
extinction-corrected, which can  introduce a considerable uncertainty.
The emissions by dust at 8~\mi\ and 24~\mi\ have the advantage of being, due to their long wavelengths, practically extinction-free SF tracers, but
they rely on the presence of dust and might therefore miss the fraction of SF which is not
attenuated.  The combinations of stellar and dust emission
(here we use 8~\mi, 24~\mi\ and \halpha/UV) are expected to be the most reliable tracers as they
are sensitive to both dust-free and dust-enshrouded
regions with no need to apply an extinction correction \citep[see][]{kennicutt12}.

We carried out aperture photometry on the FUV,  \halpha,  8~\mi\ and 24~\mi\  images 
using the aperture as  following the CO emission (total emission, NE and SW cloud, see Fig.~\ref{fig:co-mom0}). In addition, 
we derived the flux for the highest-resolution SFR tracers,  \halpha\  and  8~\mi\   using smaller
apertures centred on peaks of the 8~\mi\ emission (see Fig.~\ref{fig:irac-co}).
In order to take  the different angular resolutions of the data into account we performed a Gaussian convolution on
the  \halpha\ and  8~\mi\ map to the resolution of the taper90 CO data (4\farcs{2}).
The resolution of the GALEX FUV map is comparable to that of the taper90 CO. On the other hand, the resolution
of the  24~\mi\ is slightly worse. We ignore this effect because it is expected to be small compared to  the
flux uncertainty and because the 24~\mi\ emission is not our main SF tracer.  In this
analysis we give more weight  to the combined 8~\mi\ and \halpha\, emission which has a higher 
resolution and higher signal-to-noise. 
%Nevertheless, the comparison to the 24~\mi\ is useful 
%and keep in mind that this could lead to a slight underestimate
%of the SFR within the apertures.}

The stellar contribution to the 8~\mi\ was found to be less than 0.5\% in the CIGALE modelling,
and therefore we assume that the 8~\mi\ emission is entirely emitted by dust.
We corrected the UV emission for Galactic foreground extinction (see Sect. 3.5).
We furthermore applied a correction
for the    internal extinction based on the values derived in the 
SED modelling with CIGALE (Table~\ref{tab:output_best_fit}). The total extinction correction of the
UV flux was significant (a factor$\sim$  4).
The extinction correction of the \halpha\ flux was very small (0.07 mag, or a factor of 1.06), 
derived from the \halpha/\hbeta\ ratio 
 \citep{duc07}.

%We used different tracer for the SFR (UV,  \halpha\ , and \spitzer 8\mi\ and 24~\mi) in order to obtain a more detailed
%picture of the SF and its history. 
We used the prescriptions in Tables 1 and 2 of \citet{kennicutt12}  to
derive SFRs, both from monochromatic and  mixed (stellar+dust dust emission) tracers, except
for the calculation of the SFR from the 24~\mi\ emission where we used the relation derived  by 
%\citet{zhu08} (SFR = $8.1\, 10^{-37} L_{24\mu m}^{0.848} ({\rm erg s^{-1} }$)).
\citet{calzetti07} from resolved observations (SFR = $1.27\, 10^{-38} L_{24\mu m}^{0.848} ({\rm erg s^{-1} }$)).
In addition we used the relation from \citet{zhu08} to derive the SFR from the 8~\mi\ emission
(SFR = $1.2\, 10^{-43} L_{8\mu m} ({\rm erg\, s^{-1} }$)).
All SFRs were calculated for a Kroupa IMF \citep{kroupa01} which is very similar to the Chabrier IMF used in the
CIGALE modelling.

%%%%%%%%%%%%%%%%%%%%%%%%%%%%%%%%%%
\begin{table*}
\caption{SFR parameters for the different regions}
\begin{tabular}{lccccc}
\hline
 Band  & Flux\tablefootmark{a}& Luminosity\tablefootmark{a} & SFR\tablefootmark{b} & $\tau_{dep}$\tablefootmark{c} & $\Sigma_{\rm SFR}$\tablefootmark{b} \\
    &  [mJy]\tablefootmark{d}&  [$10^{38}$  erg s$^{-1}$] &[$10^{-4} $\msun\ yr$^{-1}$] & [$10^{10}$ yr]  & [$10^{-4}$ \msun\ yr$^{-1}$ kpc$^{-2}$] \\
    & [$10^{-15}$ erg\,cm$^{-2}$\,s$^{-1}$] \tablefootmark{d}& & & &\\ 
\hline      
{\bf Total emission} \\
\hline 
              FUV  &  0.039$\pm$  0.004  &  265.4  &   39.1  &   0.47  &   7.56 \\
         \halpha\  &   4.24$\pm$   0.42  &   1.46  &    7.8  &    2.3  &   1.52 \\
           8~\mi\  &   1.25$\pm$   0.18  & 161.33  &   19.4  &    0.9  &   3.74 \\
   8~\mi +\halpha  &         -  &         -  &   17.8  &    1.0  &   3.44 \\
          24~\mi\  &   1.00$\pm$   0.30  &  43.19  &   15.2  &    1.2  &   2.95 \\
      24~\mi +FUV  &         -  &         -  &   19.4  &    0.9  &   3.77 \\
  24~\mi +\halpha  &         -  &         -  &   12.5  &    1.5  &   2.43 \\
\hline 
{\bf NE cloud} \\ 
              FUV  &  0.030$\pm$  0.003  &  205.2  &   30.2  &   0.61  &   5.84 \\
         \halpha\  &   3.76$\pm$   0.38  &   1.22  &    6.6  &    1.7  &   2.27 \\
           8~\mi\  &   0.92$\pm$   0.12  & 118.67  &   14.2  &    0.8  &   4.94 \\
   8~\mi +\halpha  &         -  &         -  &   13.9  &    0.8  &   4.82 \\
          24~\mi\  &   0.87$\pm$   0.18  &  37.33  &   13.3  &    0.8  &   4.54 \\
      24~\mi +FUV  &         -  &         -  &   15.7  &    1.2  &   5.33 \\
\hline 
{\bf SW cloud} \\ 
              FUV  &  0.008$\pm$  0.001  &   51.7  &    7.6  &    2.4  &   1.47 \\
         \halpha\  &   0.60$\pm$   0.06  &   0.19  &    1.0  &    6.3  &   0.60 \\
           8~\mi\  &   0.26$\pm$   0.05  &  33.26  &    4.0  &    1.6  &   2.32 \\
   8~\mi +\halpha  &         -  &         -  &    3.1  &    2.1  &   1.79 \\
\hline 
\end{tabular}
\tablefoot{
\tablefoottext{a}{The fluxes and luminosities are corrected for Galactic foreground emission as explained in Sect. 3.5. }
\tablefoottext{b}{The SFRs derived from the UV flux takes into account both the  Galactic foreground extinction and an internal extinction derived from the stellar population fit (see Table 3 and Sect.  3.6).}
\tablefoottext{c}{Molecular gas depletion timescale, calculated as the ratio between the SFR from  column  4 and the molecular gas mass from Table 1.}
}
\tablefoottext{d}{The unit for most fluxes entries is mJy, expect for \halpha\ which is given in $10^{-15}$ erg\,cm$^{-2}$\,s$^{-1}$.}
\tablefoottext{e}{We did not detect any significant  emission at 24~\mi\  in the SW region. } 

\label{tab:sfr-parameters}
\end{table*}
%%%%%%%%%%%%%%%%%%%%%%%%%%%%%%%%%%

The resulting values for the SFR, together with the SFR surface densities, $\Sigma_{\rm SFR}$,
 are listed in Table~\ref{tab:sfr-parameters}. 
The uncertainty in the SFRs is only partly due to the error in the fluxes from which they are calculated, but
more by the intrinsic uncertainty in the calibration of the SFR tracers   which can be up to a factor
of 2 \citep{leroy12}.
The dust derived and combined tracers give  similar results in all regions. The SFR derived from
 the \halpha\ emission yields a SFR which is a factor $\sim 2$ lower  while the UV emission tends to
 give a higher SFR by a factor of $\sim 2 $ (both compared to dust derived and combined tracers.)
 The largest difference is thus between the SFR derived from FUV and from \halpha\ (factor 2-5).
 This result could  be due to the sensitivity of both tracers to extinction which makes them less robust
 than the combined tracers. The higher SFR derived from the UV could also be 
 the result of the decreasing SFR after a single burst of SF and the short life-time of the \halpha\ compared to the UV emitting stars.
%  due to the fact that after a single burst of star formation, the \halpha\  is short-lived compared to the FUV.
Interestingly, such an effect has also been found by \citet{lee09} for a sample of $\sim$ 300 dwarf galaxies
 that showed, in contrast to spiral galaxies, a systematically lower SFR from   \halpha\ than 
 from UV (all SFR tracers having been corrected for extinction).
 
%
%The SFR derived from the 8~\mi\ emission is a factor of 2-3 lower
% than those derived from the combined tracers.  The reason for this lower value is unclear and might just reflect 
% the uncertainty in the use of this tracer, in particular when applied to this LSB  galaxy.  
% \citet{calzetti07} discussed the use of the 8~\mi\ emission as a SF tracer
% and concluded that 8~\mi\ is more problematic  than the 24~\mi\ emission, mainly because of its dependence on metallicity and environment and due
% to the contribution of evolved stars to the heating. Even though metallicity
% is not expected to be an issue in VCC 2062, local environmental effects might play a role. 
% The use of 8~\mi\ combined with  \halpha\ or UV has been found to be more reliable  \citep{kennicutt09}.
% We therefore neglect
% the SFR derived from the 8~\mi\ emission and use the higher values derived from the other estimators.
%For our analysis,  this is a conservative approach because we avoid to overestimate the discrepancy in the SFE of \object{VCC~2062} (see below).
 
%

\subsection{Comparison to star formation laws}
\label{sec:test_sflaw}

\begin{figure}[ht!]
% \centerline{
\hspace{-0.4cm}
\includegraphics[width=10.5cm]{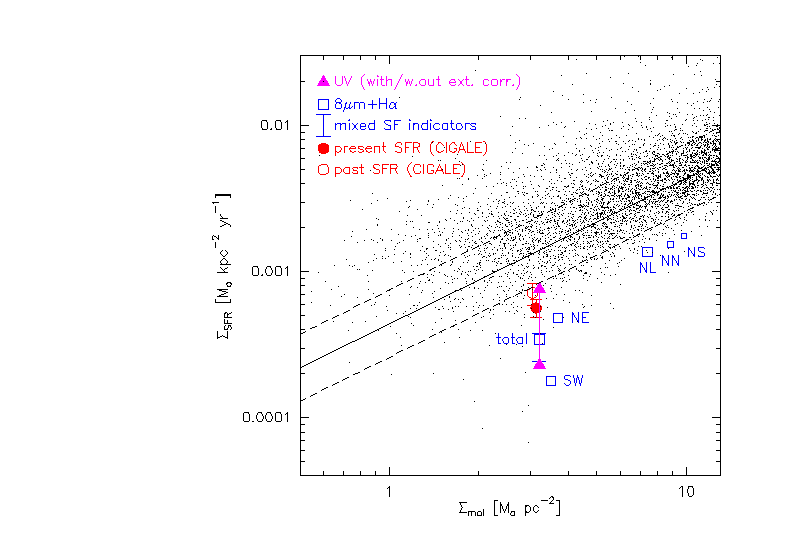}
 \caption{SFR surface density  as a function of 
the molecular gas surface density. %({\it upper left panel}).
 The values are calculated for the NE cloud, the SW cloud, the total area where CO is detected  
 (see Fig.~\ref{fig:co-mom0} for the apertures), 
 as well as for smaller apertures
 centred on the peaks of the 8~\mi\ emission in the NE cloud (see Fig.~\ref{fig:irac-co} for the apertures, labelled here 
 NL, NN and NS for the largest aperture, northern and southern apertures, respectively).
 The data of VCC~2062 compared to the results of a spatially resolved analysis of HERACLES  galaxies  \citep{bigiel11}. 
  The black dots are their data points obtained at a 1 kpc angular resolution, the
solid black line shows  their best-fit relation obtained and the dashed line its standard deviation.
}
 \label{fig:plot_schmitt_kennicutt}
\end{figure}

%We compared the surface density of the SFR, \sigmasfr, in Figs.~\ref{fig:plot_schmitt_kennicutt}a and b to the
% surface density of the gas in a classical
%Kennicutt-Schmidt  (KS) plot.
%
Fig.~\ref{fig:plot_schmitt_kennicutt} shows the comparison of \sigmasfr\ to the surface density of the CO-derived molecular gas, as the
phase that is more  closely related to SF, for different regions in VCC~2062
in a classical Kennicutt-Schmidt (KS) plot. 
We show the values derived for the different regions (NE, SW and entire CO-emitting area) and different tracers 
from Table~\ref{tab:sfr-parameters}. Furthermore we show the values for the small apertures centred on the SF peaks in the NE region.
%
%derived for each cloud and the values derived for the entire object by different tracers from Table~\ref{tab:sfr-parameters}.
We also include the present and past SFR surface density derived from the SED fitting with CIGALE (Table~\ref{tab:output_best_fit}). 
We compare the data from VCC~2062 to those of a sample of 
spatially resolved observations of nearby galaxies from HERACLES  \citep{bigiel11}\footnote{
The SFR and gas masses for this sample and the other samples used for comparison, are adapted to our choice of the IMF,
CO-to-H$_2$ conversion factor and helium fraction.}.
From these data, as well as from the
earlier studies by \citet{bigiel08} and \citet{leroy08}, a linear relation between \sigmahtwo\ and \sigmasfr\ has been 
found for angular scales between
180~pc and 1.7~kpc. 
%The quantitative analysis in these studies started at $\Sigma_{\rm H_2}$ = 5 \msun\, pc$^{-2}$
%because below the data in incomplete. For lower values in their data set (down to roughly $\Sigma_{\rm H_2} \sim$1 \msun\, pc$^{-2}$),
%the same trend is followed, but the scatter increases. 
\citet{bigiel11} compared the 
HERACLES data to an extensive data set of galaxies from the literature and found a consistent trend, albeit with an
increased  scatter (about 0.5 dex). Their linear relationship implies a constant gas depletion time of $\sim 2.35$ Gyr which can be interpreted as a 
constant SFE per molecular cloud.

\begin{figure*}[ht!]
% \centerline{
% \includegraphics[width=6.7cm]{figures/plot_schmitt_kennicutt_paper.jpg}
% \includegraphics[width=6.4cm]{figures/ks-law-kennicutt98.jpg}}
%\centerline{
% \includegraphics[width=6.7cm]{figures/krumholz12-law-kennicutt98.jpg}
% \includegraphics[width=5.65cm]{figures/tdyn-law-kennicutt98.jpg}
% \centerline{
% \includegraphics[width=6.4cm]{figures/ks-law-kennicutt98.jpg}
% \includegraphics[width=6.7cm]{figures/krumholz12-law-kennicutt98.jpg}
% \includegraphics[width=5.65cm]{figures/tdyn-law-kennicutt98.jpg}
% }
 \centerline{
  \includegraphics[width=17cm]{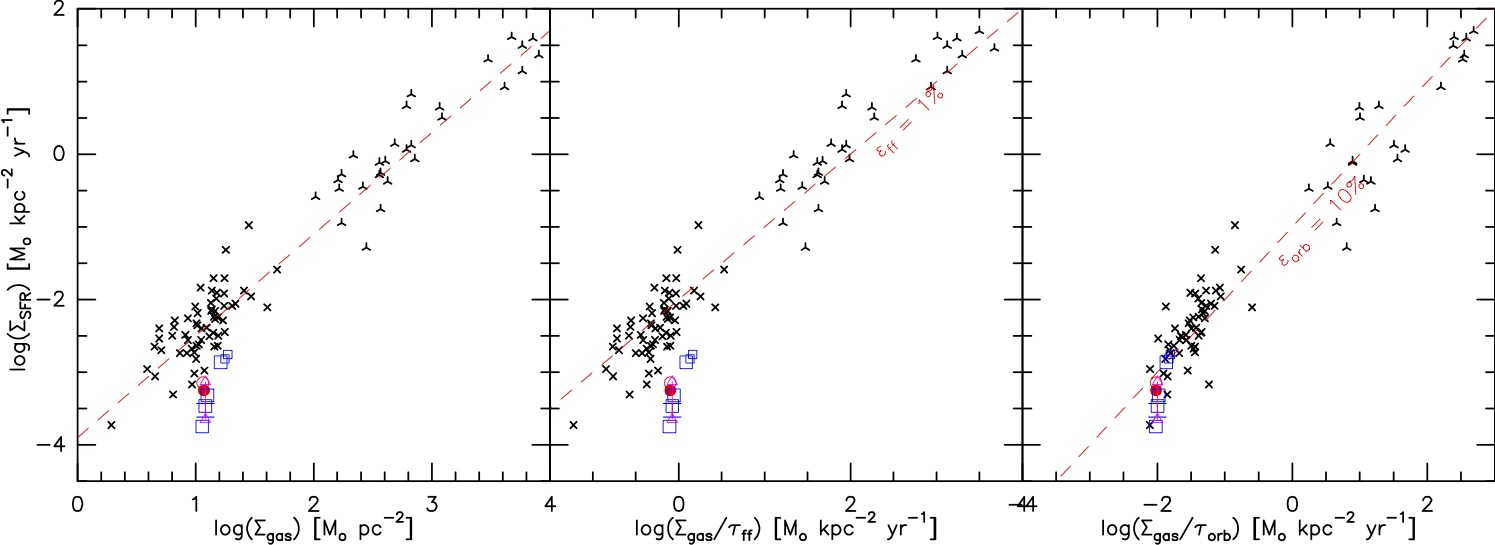}
}
%
% \includegraphics[width=9cm]{figures/tdyn-law-kennicutt98.jpg}
 %}
 \caption{SFR surface density  as a function of 
% the molecular gas surface density ({\it upper left panel}), 
the total (\htwo+HI+He) gas surface density ({\it left panel}),
 gas surface density per free-fall time ({\it middle panel}) and gas surface density per orbital timescale ({\it right panel}).
 The values for VCC~2062 (red, magenta and blue) are labelled as in
 Fig.{\ref{fig:plot_schmitt_kennicutt}}.
 %are calculated for the NE cloud, the SW cloud, the total area where CO is detected  
% (see Fig.~\ref{fig:co-mom0} for the apertures used for the integration), 
 %
%{\bf  as well as for smaller apertures
% centred on the peaks of the 8\mi\ emission in the NE cloud (see Fig.~\ref{fig:irac-co} for the apertures, labelled here 
% NL, NN and NS for the largest aperture, northern and southern apertures, respectively).}
 %
 The SFR has been calculated with different tracers (see Sect.~\ref{sect:ks-law} and Table~\ref{tab:sfr-parameters}). 
%  {\it Upper left:}  The data of \object{VCC~2062} compared to the results of a spatially resolved analysis of HERACLES  galaxies  \citep{bigiel11}. 
%  The black dots are their data points obtained at a 1 kpc angular resolution, the
%solid black line shows  their best-fit relation obtained and the dashed line its standard deviation.
{\it Left: }  The values for VCC~2062 are
compared to the sample of \citet{kennicutt98}. %The red-dashed line is the fit to his data.
{\it Middle:}  The surface density of the SFR is compared to the molecular surface density per free free-fall time, following \citet{krumholz12}.
The values for VCC~2062 are compared to the sample of \citet{kennicutt98}. 
The red dashed line gives the relation for a 1\% efficiency, the value that
\citet{krumholz12} find to describe the data well.
{\it Right: } The surface density of the SFR is compared to the molecular surface density per orbital times scale.
A value of \tauorb = 1.2 $10^9$ yr \citep{lelli15} is adopted for VCC~2062. The comparison sample is from  \citet{kennicutt98}.
The red dashed line gives the relation for a 10\% efficiency, the value that
\citet{kennicutt98} find as their average value.
%and the dashed, red
%lines the relation from Fig. 11 of Kennicutt \& Evans, adapted assuming that the molecular gas mass is about 50\%
%of the total gas mass, the value appropriate for VCC2062.
}
 \label{fig:plot_schmitt_kennicutt_compare}
\end{figure*}

In Fig.~\ref{fig:plot_schmitt_kennicutt_compare} (left) we show a similar relation, but now comparing \sigmasfr\ to the surface density of the 
total (molecular and HI) gas. We  include, for comparison, 
data for the  galaxy sample from Kennicutt (1998), spanning from spiral 
to starburst galaxies, and the corresponding
fit to those data which is a power-law in the form \sigmasfr $\propto$ \sigmagas$^{1.4}$.

As a third SF law we test the relation proposed by \citet{krumholz12} (Fig.~\ref{fig:plot_schmitt_kennicutt_compare} (middle). 
They derived a good relation between \sigmasfr\ and \sigmagas/\tauff, 
holding from molecular cloud scales to starburst galaxies, where
\tauff\ is the local free-fall time of the gas. This quantity can be calculated, assuming that   the molecular gas is distributed in 
GMCs, as
\begin{equation}
\tau_{\rm ff} = \frac{\pi^{1/4}}{\sqrt{8}}\left[\frac{\sigma}{G\Sigma_{\rm GMC}^3\Sigma_{\rm gas}}\right]^{1/4},
\end{equation}
where $G$ is the gravitational constant, $\sigma$ the gas velocity dispersion, $\Sigma_{\rm GMC}$ the surface density of a typical molecular
cloud, adopted, following \citet{krumholz12},  as  85 \msun\ pc$^{-2}$, and  $\Sigma_{\rm gas}$ the average gas surface density in the
region where the stars form. We adopt for the molecular gas the same velocity dispersion as for the atomic gas
\citep[$\sigma = 7$ \kms , ][]{lelli15}. We note, however, that this model should be applied with caution to VCC~2062 where
not only gravitational but also tidal forces might be important.

The values  of $\Sigma_{\rm SFR}$ of VCC 2062 
lie below all these three relations. The difference is particularly large for the SFR
derived from the hybrid SF indicators, and  largest for the  SW cloud (more than a factor of 10).
The difference for the SFRs derived from the extinction-corrected
UV emission and with CIGALE  are less pronounced.  They  are on the low side, 
just outside the scatter, of what is expected from the spiral galaxy sample.
The values of both \sigmasfr\ and \sigmamol\ depend on the size of the aperture and increase for smaller apertures which are
 centred on peaks of the SF region. However, 
they still  lie well below the three relations even for the smaller apertures.

As a final test, we  looked into the relation between \sigmasfr\  and \sigmagas/\tauorb\
\citep{elmegreen97,silk97}.
\citet{kennicutt98} found that this prescription 
adjusted the data as well as just with   \sigmagas. 
Moreover, starburst galaxies at high redshifts seem to deviate from the standard KS relation but adhere to its kinematical version with orbital 
times \citep{daddi10,genzel10}.
This relation implies that per rotation period about 10\% of the
available gas is converted into stars. 
The physical reason for this relation is still open and includes processes such as
 spiral density waves  or, in a slightly modified model,  cloud-cloud collisions \citep{tan00}
 or cloud collapse due to a large-scale Toomre instability \citep{krumholz12}.
 Intriguingly, a similar relation between \sigmasfr\  and  \sigmagas/\tauorb\ also holds for dwarf galaxies, which lack a well-defined spiral 
 pattern and are probably forming stars in a HI-dominated regime \citep{lelli14}.
There are, however, also  observations that do not strongly support  this prescription.
 \citet{leroy08} compared the orbital timescale to the radial
variation of the SFE and concluded that \tauorb\ alone cannot explain SF, and
 \citet{krumholz12} showed that a relation between \sigmasfr\  and \sigmagas/\tauorb\ is not able to
explain the process of SF from very small (molecular clouds) to large (starburst) scales.
%
%The model by \citet{krumholz12}  which proposed the free-fall times as parameter determining the SFR,
%also predicts a  dependence  on the orbital time for starburst galaxies because their
%ISM is not formed by individual molecular clouds but by a dense ISM whose global stability is governed by a Toomre
%criterion. 

Due to the lack of  DM,   VCC~2062 has a long orbital period of, derived from the HI,
$1.2 \times 10^9$ yr (Lelli et al. 2015).  Together with the mean gas 
depletion time of $\sim 1-2 \times 10^{10}$~yr (derived from the mixed tracers, see Table 4)
we derive a gas consumption time per revolution of 6-12\% in the range of the value of 10\%
found by  \citet{kennicutt98}
 In Fig.~\ref{fig:plot_schmitt_kennicutt_compare} (right) we compare the relation derived with this value
for VCC~2062 to that of the sample from \citet{kennicutt98}. Indeed,
VCC~2062 follows this relation well, with a very good agreement
for the NE region and the total emission and an offset towards low values only for the SE region.
%Interestingly, the ratio between this
%orbital time and the mean gas 
%depletion time of $\sim 1-2 \times 10^{10}$~yr (derived from the mixed tracers, see Table 4)
%%, the values derived from the extinction-corrected UV and from the SFR from CIGALE are shorter,
%is about 0.06 and gives  a  
%gas consumption per revolution of 6-12\%,
%very close to the value of 10\% 
%found by  \citet{kennicutt98}. % for different galaxies, from normal to starbursts.
%These values are in the same range as derived for \object{VCC~2062} from extinction-corrected UV and similar to those
%derived from the SFR from CIGALE (\taudep = 3-5 Gyr).
%
The similarity of the SFE per revolution in  VCC 2062 and spiral/starburst  galaxies could indicate that
galaxy-wide processes related to the rotation are responsible for the collapse of molecular clouds.
%If this were the case,  the
%apparently low SFE derived from a KS-law (Fig.~\ref{fig:plot_schmitt_kennicutt}a+b) might be misleading and due to the fact that the (unusually long) orbital time
%in the DM free \object{VCC~2062} has not been  taken into account. 
If this were the reason we would expect a similar result for TDGs in general.
For the TDG in Arp ~158 \citep{boquien11}, the two TDGs in Stephan's Quintet \citep{lisenfeld04}
and a small sample of TDGs detected in CO  \citep{braine01}, \taudep\ could be estimated 
and was found -- contrary to VCC~2062 -- to be similar to that of spiral galaxies ($1-3 \times 10^9$~yr).
These observations are mostly based on single pointings with the IRAM 30m telescope which do not
resolve the structure of the TDGs and therefore do not allow us to apply
kinematical models to derive  \tauorb.
%
%In a rough estimate, we  can use  the unresolved CO data to calculate an upper limit of
 %the orbital time scale as \tauorb = $ 2\pi R/v$, where we approximated the    radius by 
%half the beam size (which might overestimate the real radius), and the rotational velocity, $v$, as  half the CO line-width.  
%The orbital times scales for the 7 sources with reliable CO linewidths \citep[see Table 5 in ][]{braine01} lies between         
%1 and 4 $\times 10^9$ yr, with a mean value of $2\times 10^9$ yr, i.e. similar to \taudep.
Thus, the analysis of  the  available data of other TDGs does  not allow to draw any firm conclusions
with respect to the role
played by \tauorb\ in the SF process.

%%%%%%%%%%%%%%%%%%%%%%%%%%%%%%%%%%

%%%%%%%%%%%%%%%%%%%%%%%%%%%%%%%%%%
%%%%%%%%%%%%%%%%%%%%%%%%%%%%%%%%%%

\subsection{The specific star formation rate}

We used the present SFR and the stellar mass derived from the stellar population CIGALE  fitting to 
calculate the specific SFR (sSFR), as sSFR = SFR/M$_*$, yielding  sSFR = $5.5 \times 10^{-10}$ yr$^{-1}$.
%It is unclear what fraction of the stars in this area are really associated with \object{VCC~2062} and which
%fraction are from the parent galaxy and have not formed in situ, as suggested by the bridge of olds stars
%visible in Fig.~\ref{fig:ngc4694-deep-image}. 
This value lies in the range typical for  low-mass 
spiral galaxies  with stellar masses of  $10^8-10^9$ \msun\ \citep[see][for a summary]{kennicutt12}. 
Applying the relation from \citet{Schiminovich07} (log(sSFR) = -0.36 log(M$_*$)-6.4) 
 we would expect a somewhat higher sSFR for the
low-mass VCC 2062 (M$_* = 7\, 10^6$ \msun), of  the order of $10^{-9}$ yr$^{-1}$.
The most likely reason for the somewhat lower value is that VCC~2062 is a recycled galaxy so that
part of its stars might come from the parent galaxy and have not formed {\it in situ}. Fig.~\ref{fig:ngc4694-deep-image} 
supports this conclusion since it clearly shows a 
bridge of old stars from NGC 4695 to VCC 2062. 
We can test this by deriving the sSFR only in the part of the galaxy where recent SF takes place, assuming that all
the stars here have formed in VCC~2062.  When we use  the CIGALE value for the SFR and M$_*$ for
the CO-emitting area and
derive sSFR = $9 \times 10^{-10}$ yr$^{-1}$, close to the expected value.

%__________________________________________________________________

\section{Discussion}

\subsection{VCC~2062 as an old TDG}

\citet{duc07} studied VCC 2062 in detail and concluded that it is most likely an old TDG.
In this study we confirm this view. Two important new pieces of evidence reinforce  the past conclusion.
(i) A kinematical analysis by \citet{lelli15} shows that there is no evidence for DM in VCC 2062, 
as predicted for this class of objects. (ii) The deep NGVS image reveals a bridge between NGC 4694 and 
VCC 2062 (see Fig.~\ref{fig:ngc4694-deep-image}), clearly showing the connection between these objects.
Besides, the new deep image shows that the putative parent galaxy exhibits many tidal disturbances 
consistent with the hypothesis that it is the result from a major gas-rich merger, i.e. the type of collisions able to form TDGs.

The large, multiwavelength data set available for VCC 2062 has allowed detailed analysis of the present and past SF,
making  VCC 2062 the TDG with the best-studied SED. 
Previous, similiar studies of TDGs and intergalactic SF regions \citep{boquien07, boquien09, boquien10}, had to base their
SED modelling on a more incomplete wavelength range which limited the conclusions, for example about the old stellar population.
Our analysis shows that the
SFR has been declining in the past with a burst that took place about $\sim$ 300 Myr ago and  included 
a considerable fraction of the mass of the object. 
This timescale is shorter that the typical merger timescale of $\sim$ 1 Gyr \citep{bournaud10}, consistent with the
picture that the SF has taken place in the TDG.

The mass fraction of stars formed in the burst ($\sim 40\%$) is higher than typical values in spiral and 
dwarf galaxies \citep[see][finding that the burst strengths is $<$ 1\% in Blue Compact Dwarf Galaxies]{depaz03} and typical of what is found in TDGs
for which in some cases no  old stellar population is found  at all  \citep{boquien09,boquien10}.
An open question that remains is which fraction of the old stellar population is  actually associated
with the gravitationally bound object, VCC~2062, and which part belongs to the tidal tail, visible in Fig.~\ref{fig:ngc4694-deep-image}.
About half of the stellar mass derived for the entire object is situated in the CO-emitting region with
young SF. Most likely, at least this fraction of the stellar mass belongs to VCC~2062. However, in order
to definitively conclude this, deep spectroscopical observations determining the kinematical association
of the old stars with VCC~2062 or the tail would be necessary.

\subsection{Comparison to  other TDGs}

%%%%%%%%%%%%%%%%%%%%%%%%%%%%%%%%%%
\begin{figure}[ht!]
 \centerline{
 \includegraphics[width=8cm]{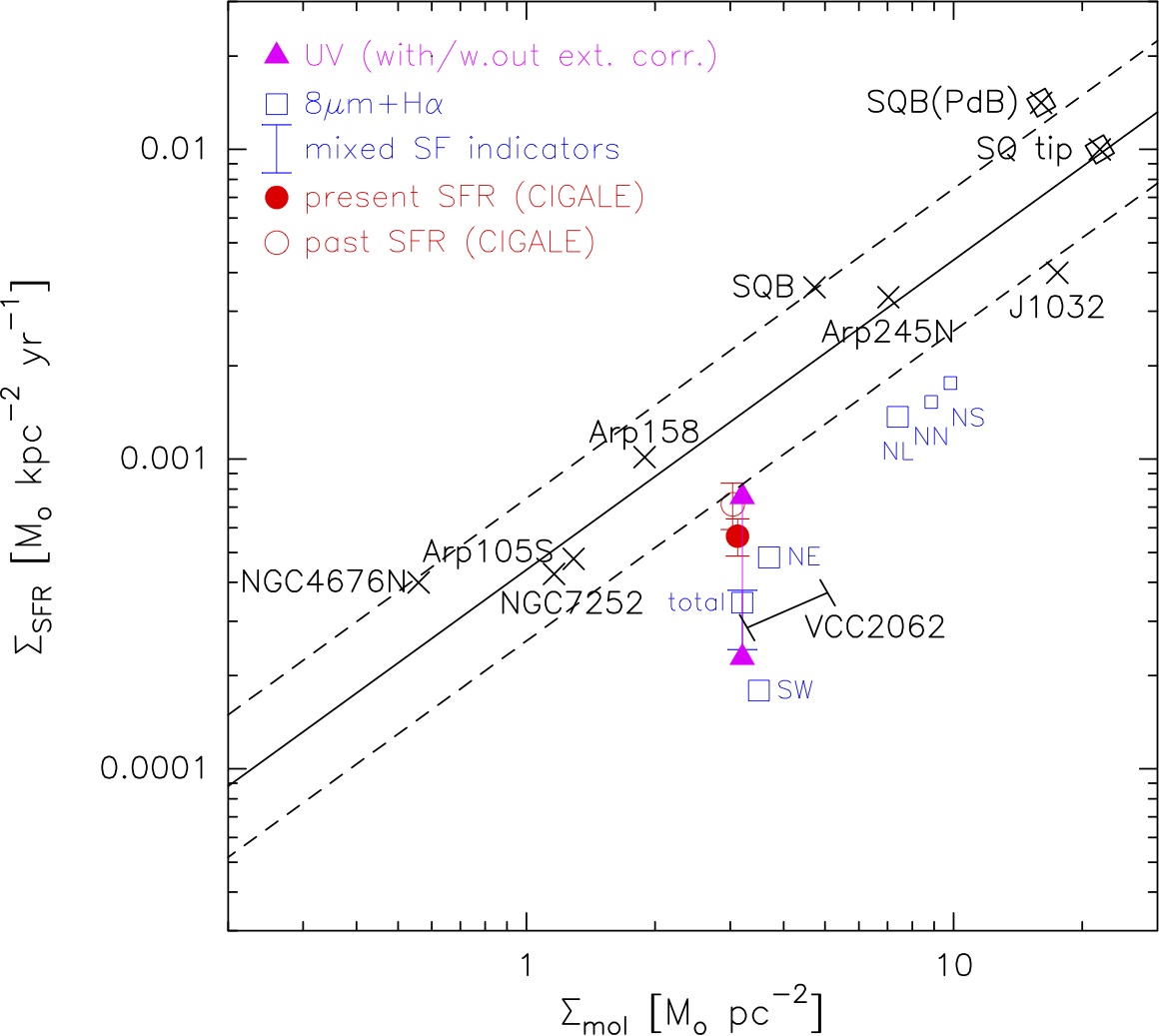}
 }
\caption{
Molecular Kennicutt-Schmidt (KS) relation of a small sample of TDGs
for which data for molecular gas and the SFR is available in the literature
(see Table~\ref{tab:comp_tdg} for the values and references).
Most observations are carried out with a single-dish telescope (IRAM 30m telescope) except for
two objects in Stephan's Quintet (SQ) which are marked with a square symbol.
We also include the values for VCC~2062 from IRAM 30m data (black short line), both for the 
entire galaxy (lower end of the line) and for a single pointing centred on the
NE region (upper end of the line).
The \citet{bigiel11} relation for spiral galaxies and the PdBI values for VCC~2062 
(in red, magenta and blue) are  shown.
%, see Fig.~\ref{fig:plot_schmitt_kennicutt} (left) for the coding.}
}
 \label{fig:plot_ks_tdg}
\end{figure}

\begin{table}
\caption{Data for the molecular gas and SFR for a sample of TDGs}
\resizebox{\linewidth}{!}{%
\begin{tabular}{lllllll}
\hline
%\small{
Name & Dist & FWHM  & Area & \mmol & SFR & Ref.\\ 
    & [Mpc] & [\arcsec/kpc] & [kpc$^2$] &  [$10^7$  & [$10^{-2}$   &\\
    &            &    &   &   \msun] & \msun yr$^{-1}$] \\
\hline
    \object{Arp158}   &   62.1   &   22   /    6.6   &   49.3   &    8.5   &   5.0 & (1)  \\
    \object{Arp105S }   &  115.0   &   22   /   12.2   &  169.9   &   19.8   &   8.1  & (2)\\
   \object{NGC4676N}    &   90.0   &   22   /    9.6   &  201.5   &   10.3   &   8.1 & (2) \\
   \object{NGC7252W}    &   64.0   &   11   /    3.3   &   12.6   &    1.3   &   0.5 & (2) \\
%  NGC5291N   &   58.0   &   45   /   12.6   &  179.8   &   21.1   &   4.78 & (2) \\
%  NGC5291S   &   58.0   &   45   /   12.6   &  179.8   &   28.2   &   2.36  & (2)\\
        \object{SQB}    &   85.0   &   22   /   10.0   &  112.2   &   48.3   &  32.2  & (2,3)\\
        SQB   &   85.0   &    4   /    1.6   &   21.2   &  34.0   &  30.0 & (4) \\
      \object{SQtip}    &   85.0   &    4   /    1.6   &    5.0   &  11.0   &   5.0  & (4)\\
 \object{J1032+1952}    &   20.4   &   22   /    2.2   &    6.0   &  10.5   &   2.4  & (5) \\
    \object{Arp245N}    &   31.0   &   22   /    3.3   &   27.0   &  19.0   &   9.0  & (6,7)\\
    VCC~2062  (NE) & 17 & 22 / 1.8 &  7   & 2.3   & 0.2 & (8,9) \\
    VCC~2062  (total) & 17 & 22 / 1.8 &  2.7  & 1.4   & 0.2 & (8,9) \\
   \hline
\end{tabular}
}
\tablefoot{
All data are adapted to include He in the molecular gas mass and to a Kroupa IMF.
The SFR is mostly derived from \halpha\, except for Arp~158 where it is derived
from FUV and 24~\mi. The area denotes the area over which the \mmol\ and SFR have
been determined which can be larger than the resolution of the observation.
References for the molecular gas and SFR dat.  
(1) \citet{boquien11} , 
(2) \citet{braine01} (Table 2) and reference therein, 
(3) \citet{lisenfeld02a},
(4) \citet{lisenfeld04},
(5) \citet{lisenfeld08},
(6) \citet{braine00},
(7) \citet{duc00},
(8) \citet{duc07} (for the CO from IRAM 30m),
(9) this paper (for the SFR).
\label{tab:comp_tdg}
}
\end{table}

In Fig.~\ref{fig:plot_ks_tdg} we show the KS-relation for a small sample
of TDGs taken from the literature (see Table~\ref{tab:comp_tdg}),  together with
the data for VCC~2062.
%
%where VCC~2062 falls on the KS-relation,
%together with similar
%data for other TDGs taken from the literature
%(see Table~\ref{tab:comp_tdg}). Most of the  other objects follow the
%standard KS-relation.
Most of the literature data have been obtained with single dish
telescopes (IRAM 30m) and  have therefore a poor angular resolution, corresponding to 
linear scales up to 12.2~kpc (see Table~\ref{tab:comp_tdg}),
much larger than the size of the SF regions.
This has two consequences. (i) The surface densities are unrealistically low because
the molecular gas mass and SFR are averaged over a large area. (ii)
There is some uncertainty as to where on the KS-relation the objects would
fall for smaller apertures centred on the SF regions.

The fact that almost all TDGs follow the standard KS-relation indicates that
SF in TDGs seems to proceed in the same way as in spiral galaxies  in spite of the very different
local conditions (DM content, presence of spiral density waves and tidal forces) 
in both types of objects.
The noticeable exception is VCC~2062 which lies below the standard KS-relation.

\subsection{Why is the SFE low in VCC 2062?}

The SFR surface density in VCC 2062 is  below what is expected from its
molecular gas surface density, both when comparing it to spiral galaxies or other TDGs.
The  difference is  particularly large for the SW region
which has only weak SF but a considerable amount of molecular gas.
The difference is larger for SFRs based on dust and \halpha\  than for those based on UV or CIGALE modelling.
Furthermore, the  difference decreases for smaller apertures which are centred on the 
peaks of SF visible in 8~\mi\ within the NE region, but is still there. Thus, even though the exact value
of the SFE varies with the choice of SF tracer and with the angular resolution, we find the consistent
result that the SFE is low in VCC~2062 compared to other TDGs and spiral galaxies.

The molecular gas surface density in VCC 2062 is very low, among the lowest probed in similar studies
\citep[e.g.][]{bigiel08,bigiel11,schruba11}.   
%Could this low surface brightness/molecular gas surface density be the reason
%for the low SFE?
%
SF in low gas surface densities has been studied  in the outskirts of galaxies \citep{schruba11} or in LSB galaxies \citep{boissier08,wyder09} 
and very low  values of SFE(gas)= SFR/\mgas\  have been found.
However, in these regions 
molecular gas formation is not very efficient and the molecular gas fraction is low.
%\citet{wyder09} studied the SF law in LSB galaxies based on the atomic gas (since no molecular gas data was available
%for most of their objects). They found that the SFR was below the expectation of the SF law extrapolated 
%from higher surface brightness galaxies. They compared their results to models of \citet{krumholz05} and concluded that the low SFE(gas)
%was consistent  with a decreasing formation of molecular gas from atomic gas at these low surface brightnesses.
Possibly, the SFE based on the (undetected) molecular gas would have been normal. This is indeed what has been
found in the outskirts of galaxies by stacking the CO spectra
\citet{schruba11}.
%
%studied the SF law in disks of spiral galaxies and found a very low  SFE(gas) in the galactic outskirts where the molecular gas was undetected.
%However, by stacking the molecular gas spectra in these regions they showed  that the   SFE (per molecular gas)
%was roughly constant over the disk, independent of whether the gas is molecular (in the inner disk) or atomic dominated (in the outskirts).
These results are consistent with the picture that at low surface brightnesses the formation of molecular gas is inefficient, but when it
is present, it forms stars with the same SFE as in the inner disks of spirals.
The situation in the LSB galaxy  VCC~2062 is very different from the objects described above. Here, 
the molecular gas fraction is considerable (\mmol/\mhi $\approx 0.3$). 
The question is why this molecular gas is  forming stars at a very low efficiency.

VCC~2062 is not the only object with abundant molecular gas but little SF.  Similar situations have been observed in 
   other unusual objects, for example in some regions of the TDG candidate J1023+1952 in the system Arp 94 \citep{lisenfeld08},
the intergalactic bridge in Stephan's Quintet \citep{lisenfeld02a, guillard12}, the face-on colliding  Taffy galaxies
\citep{braine03, braine04}, some galaxies in Hickson Compact Groups \citep[HCG;][]{alatalo15} or in the gas in tails stripped from
galaxies in clusters \citep{jachym14,verdugo15}.
The reason for the lack of SF in spite of the presence of abundant, CO-traced molecular gas  is not completely clear, and might be different in
each case. A possible reason is the lack of a dense  molecular gas phase, as in the bridge of the Taffy galaxies.  \citet{braine03} showed
for this system
that molecular gas reformed  quickly after the collision, but the dense  molecular cores were destroyed. In galaxies in HCGs, 
\citet{alatalo15} argued that shock-injected turbulence perturbed the molecular gas and
decreased the SFE.

The situation in VCC~2062 is probably different from the above examples. It is most likely a self-gravitating object, similar to
other TDGs in which, in general, 
the SFE seems  normal (see Sect.~\ref{sec:test_sflaw}).
However, the much lower surface brightness of VCC~2062,
both with respect to the gas and to the stars, is a major difference between VCC~2062 and other studied TDGs.
The stellar surface density (0.6 \msun\ pc$^{-2}$ within the CO-emitting area) is considerably lower than  \sigmagas\  which 
may lead to a flaring in the gas disk, similar to what is seen in the outer  disks of spirals and irregular galaxies \citep{elmegreen15a},
causing a decrease in the  volume density of the gas.
%\citet{elmegreen15b} included this effect and 
%predicted for this regime a relation \sigmasfr = $1.5\ 10^{-5}$ (\sigmagas$)^2$ (adopting a velocity dispersion of 7~\kms)
%which yields  for \object{VCC~2062}
%(\sigmagas $\sim$ 12  \msun\ pc$^{-2}$) a value of  \sigmasfr = $2 \, 10^{-3} $ \msun\ yr$^{-1}$ kpc$^{-2}$, a factor
%between 3 and more than 10 higher than what is observed. 
%Thus, even though the disk scale height in \object{VCC~2062} is most likely large, this  effect by itself does
%not seem to explain entirely the low SFR.
An additional effect  of the low surface density of newly formed stars is that their combined radiation pressure is less efficient in 
dispersing  the molecular gas.  Thus, we might simply be seeing the remaining molecular
gas, left-over after the formation of stars.  
Due to the low stellar surface brightness the interstellar radiation field and thus the gas temperature are  most likely low.  
\citet{shetty11} showed that there is only a weak dependence of \xco\ on the gas temperature, $T$, (\xco\  $\propto  T^{-0.5}$ for $T \sim 20-100$ K).
Apart from being weak, the direction of this dependence means that we might slightly {\em underestimate} the molecular gas mass,
and thus {\em overestimate} the SFE, i.e. making the difference with respect to spiral galaxies even larger.

The kinematic information of the gas allows us to apply the stability criterium of \citet{toomre64} and derive the 
critical gas surface density for SF as 

\begin{equation}
\Sigma_{\rm crit, gas}  =  \frac{\alpha \sigma \kappa}{\pi G},
\end{equation}
where $\alpha$ is a constant which we adopt, following \citet{martin01}, as $\alpha = 0.69 $ and $\kappa$ is the epicyclic
frequency.  We derived critical densities of 2 and 4 \msun\ pc$^{-2}$ for the SW and NE regions, albeit with a high uncertainly
(factor 2-3), mostly because of the 
uncertainty in the rotation velocity which is a factor of 2 \citep[see Tab. 7 in][]{lelli15}.
However, even considering this uncertainty, the observed gas surface densities (between 11 and 13 \msun\ pc$^{-2}$, see Tab.~\ref{tab:flux_mass_regions})
seem to  be above the estimated critical value so that the gas is expected to be able to form stars.

%\begin{equation}
%\kappa = 1.41  \frac{v}{r} \sqrt{1+\beta},
%\end{equation}
%where $\beta = d\log(v)/d\log(r)$. 

Finally, the fact that VCC~2062 is a small object where SF is intermittent might play a role when comparing its SF law
to that of larger spiral galaxies where local variations average out more easily. In fact, the analysis of the HI
distribution and kinematics \citep{lelli15} has shown that the SF regions are distributed in a very
asymmetric way in VCC~2062 (see Fig.~\ref{fig:pv-hi-co-lelli}). The young SF regions  all lie on one side of the kinematical
centre, whereas the old stars, traced by optical emission,  are found also on the other side
and extend to the tidal tail connecting VCC 2062 to the parent galaxy.
Temporal variations in the SF history  could have two consequences. 
(i) The calibration of the SF tracers is different
from that in spiral galaxies. %as we have seen in the difference between UV, dust and \halpha\ tracers.
Indeed, \citet{leroy12}
has shown that SF tracers are not reliable and do not give very consistent results among themselves on kpc scales and  for 
values of \sigmasfr $\lesssim 10^{-2}$ \msun\ yr$^{-1}$ kpc$^{-1}$. %, which is the regime of \object{VCC~2062}.
%However, we are
%ntegrating over areas on a $\ga$ 1kpc scale  and thus most likely average over different  SF regions. 
%On the other hand, the fact that the SFR derived from CIGALE is higher than that derived from 
%most of the other SFR tracers, indeed casts doubt on the reliability of the former.
(ii) Our observations might
be sensitive to short temporal variations in the ratio between SFR tracers and molecular gas mass which would
average out in larger spiral galaxies. Such an effect is expected to be particularly relevant on small angular scales
as in  the small and faint region SE where the
SFE is indeed particularly low.

In summary, there are a number of reasons why the SFE in VCC~2062 is  low. 
We cannot firmly conclude which are the most relevant ones, and very likely, several of them play a role.
In any case, it is very likely that the low 
surface brightness of the object (both in stars and gas) plays an important role
in lowering the gas density, decreasing the stellar radiation feedback and increasing the
uncertainties in the standard SFR tracers.

%__________________________________________________________________

\section{Summary and conclusions}

We presented new CO(1-0) data  obtained with the Plateau de Bure Interferometer of VCC 2062,
a nearby TDG at the outskirts of the Virgo Cluster.
These data have allowed us to study the molecular gas distribution and kinematics. Using a large set
of complementary  data we  modelled the UV-to-IR SED and derived the SFR from different tracers.
Combining the SFR with  the CO data, we studied the SF law 
in this object.  TDGs have two important properties that make this analysis particularly interesting:
(i) In contrast to classical dwarf galaxies, their metallicity is higher (typically of the order of 0.5-0.3 of the Milky Way value) 
and  therefore CO is a good tracer of the
molecular gas, and (ii) their DM content is very low and so  we are able to study SF in this 
extreme, DM-free environment. The main conclusions are summarised as follows.

\begin{itemize}

\item The CO emission is distributed in two main regions, called NE and SW. The 
NE region coincides with the peak of the atomic gas and with SF tracers, whereas the SW region
lies in between two HI peaks and has little associated SF.

\item 
A comparison of the atomic and molecular gas showed that the atomic gas is more
extended than the CO and young SF regions.  CO is only found on one side of 
the kinematical centre of the galaxy.
The HI and CO kinematics agree well in the regions where their emissions coincide.

\item New, deep optical observations of this system with the NGVS revealed a stellar bridge
between the parent galaxy NGC 4695 and VCC 2062. This, together with the negligible dark matter
content found from HI kinematical modelling (Lelli et al. 2015), gives the final proof that VCC 2062 is a 
TDG.

\item We modelled the UV-to-IR SED with CIGALE and found that the SFR has noticeably  declined in the recent
past (a factor of 1.3-1.4 between the present SFR and the SFR averaged over the past $10^8$ yr).

\item We calculated the SFR from various tracers (UV, \halpha, 8,  and 24 \mi\ and combinations of them).
In general the different tracers gave compatible results with the exception of the UV which gave a factor $\sim$ 2-4
higher SFR. A declining SFR or a paucity of massive stars are possible reasons.
The SFR surface density derived from the dust emission and \halpha\ was lower than those derived by the SED modelling
from CIGALE by a factor of $\sim$ 2-3.

\item
The molecular gas depletion time  (= \mmol/SFR),
derived at different angular resolutions between  0.3 and $\sim$ 1.5 kpc, of about $1-2\times 10^{10}$  Gyr, derived from \halpha\ and
dust tracers, is long
compared to the median value found by \citet{bigiel11} (\taudep = 2.35 Gyr,   for different samples of galaxies).
%(a factor  of 8 higher, 3$\sigma$ from the median). 
The discrepancy is particularly large for the SW region 
(gas depletion time of $\sim 6\times 10^{10}$ yr).
Similar discrepancies were found   when comparing VCC~2062 to other SF laws
\citep{kennicutt98, krumholz12}.
%The difference is less for the \taudep\ derived from the CIGALE modelling. 
%We have not found convincing evidence for an influence of the rotation period which is relatively
%low in this DM-free object.
%
%\item

\item We compare the SFE of VCC~2062 to a small sample of other TDGs, most of them with
data for the molecular gas at a poorer angular resolution.  We find that TDGs in general follow the same 
KS-relation as spiral galaxies and that VCC~2062 is the only exception.

\item We discussed various possible reasons for this  long molecular gas depletion time.
The difference is {\it not} due to an inefficient formation of the molecular gas since  
abundant molecular gas is present.
The  low surface brightness of  VCC~2062 most likely plays an important role in changing
the physical conditions compared to more massive objects (decreased mid-plane pressure, low
stellar radiation field).  Temporal effects due to intermittent SF and uncertainties in the calibration of the SF tracer
might also play a role.

\item Together with the orbital timescale \tauorb\ derived from the HI analysis \citep[$1.2 \times 10^{9}$ yr,][]{lelli15},
we derived a gas consumption per orbit of 6-12\%, in the range of  the value of 10\% derived for spiral a starburst
galaxies in spite of the lack of DM in VCC~2062. High-resolution data for other TDGs are necessary in order to find out whether this result holds for
TDGs in general which would indicate  an important role of  \tauorb\ in the SF process.

%However,  data for a small sample of  7 TDGs do not confirm this trend, but show gas depletion and
%orbital time scale are similar. Thus, the orbital times scale does not seem to play a relevant part
%in the SF law in TDGs.

\end{itemize}

\acknowledgements

We would like to warmly thank our support astronomer at Plateau de Bure, A. Castro-Carrizo,
for dedicated help during the data reduction, and F. Bigiel for making his data available to us.
We appreciate the referee's very useful comments  which helped derive firmer conclusions from the
data. 
This project made use of data obtained as part of the CFHT NGVS Large Programme. The data were processed by Jean-Charles Cuillandre and  
Stephen Gwyn under the supervision of Laura Ferrarese and Patrick C™tŽ. They are warmly thanked. 
UL acknowledges support by the research
projects AYA2011-24728 and AYA2014-53506-P  financed by the Spanish Ministerio de Econom\'ia
y Competividad and by FEDER (Fondo Europeo de Desarrollo Regional) and the Junta de Andaluc\'ia (Spain)
grants FQM108. 
EB acknowledges support from the UK's Science and Technology Facilities Council (grant numbers ST/J001333/1, ST/M001008/1).
This work is based on observations carried out under project number ue1e-2010 with the IRAM Plateau de Bure Interferometer. 
IRAM is supported by INSU/CNRS (France), MPG (Germany) and IGN (Spain).
This research has made use of the NASA/IPAC Extragalactic Database (NED) which is operated by the Jet Propulsion Laboratory, California Institute of Technology, under contract with the National Aeronautics and Space Administration. 
We acknowledge the use of the HyperLeda database (http://leda.univ-lyon1.fr).

%webpage for extinction calculator http://dogwood.physics.mcmaster.ca/Acurve.html

%%%%%%%%%%%%%%%%%%%%%%
%%%% BIBLIOGRAPHY %%%%
%%%%%%%%%%%%%%%%%%%%%%

\bibliographystyle{aa}
\bibliography{bib_vcc2062}

\begin{thebibliography}{79}
\expandafter\ifx\csname natexlab\endcsname\relax\def\natexlab#1{#1}\fi

\bibitem[{{Alatalo} {et~al.}(2015){Alatalo}, {Appleton}, {Lisenfeld},
  {Bitsakis}, {Lanz}, {Lacy}, {Charmandaris}, {Cluver}, {Dopita}, {Guillard},
  {Jarrett}, {Kewley}, {Nyland}, {Ogle}, {Rasmussen}, {Rich},
  {Verdes-Montenegro}, {Xu}, \& {Yun}}]{alatalo15}
{Alatalo}, K., {Appleton}, P.~N., {Lisenfeld}, U., {et~al.} 2015, ArXiv
  e-prints

\bibitem[{{Asplund} {et~al.}(2005){Asplund}, {Grevesse}, \&
  {Sauval}}]{asplund05}
{Asplund}, M., {Grevesse}, N., \& {Sauval}, A.~J. 2005, in Astronomical Society
  of the Pacific Conference Series, Vol. 336, Cosmic Abundances as Records of
  Stellar Evolution and Nucleosynthesis, ed. T.~G. {Barnes}, III \& F.~N.
  {Bash}, 25

\bibitem[{{Bigiel} {et~al.}(2008){Bigiel}, {Leroy}, {Walter}, {Brinks}, {de
  Blok}, {Madore}, \& {Thornley}}]{bigiel08}
{Bigiel}, F., {Leroy}, A., {Walter}, F., {et~al.} 2008, \aj, 136, 2846

\bibitem[{{Bigiel} {et~al.}(2011){Bigiel}, {Leroy}, {Walter}, {Brinks}, {de
  Blok}, {Kramer}, {Rix}, {Schruba}, {Schuster}, {Usero}, \&
  {Wiesemeyer}}]{bigiel11}
{Bigiel}, F., {Leroy}, A.~K., {Walter}, F., {et~al.} 2011, \apjl, 730, L13

\bibitem[{{Boissier} {et~al.}(2008){Boissier}, {Gil de Paz}, {Boselli}, {Buat},
  {Madore}, {Chemin}, {Balkowski}, {Amram}, {Carignan}, \& {van
  Driel}}]{boissier08}
{Boissier}, S., {Gil de Paz}, A., {Boselli}, A., {et~al.} 2008, \apj, 681, 244

\bibitem[{{Bolatto} {et~al.}(2013){Bolatto}, {Wolfire}, \& {Leroy}}]{bolatto13}
{Bolatto}, A.~D., {Wolfire}, M., \& {Leroy}, A.~K. 2013, \araa, 51, 207

\bibitem[{{Boquien} {et~al.}(2007){Boquien}, {Duc}, {Braine}, {Brinks},
  {Lisenfeld}, \& {Charmandaris}}]{boquien07}
{Boquien}, M., {Duc}, P.-A., {Braine}, J., {et~al.} 2007, \aap, 467, 93

\bibitem[{{Boquien} {et~al.}(2010){Boquien}, {Duc}, {Galliano}, {Braine},
  {Lisenfeld}, {Charmandaris}, \& {Appleton}}]{boquien10}
{Boquien}, M., {Duc}, P.-A., {Galliano}, F., {et~al.} 2010, \aj, 140, 2124

\bibitem[{{Boquien} {et~al.}(2009){Boquien}, {Duc}, {Wu}, {Charmandaris},
  {Lisenfeld}, {Braine}, {Brinks}, {Iglesias-P{\'a}ramo}, \& {Xu}}]{boquien09}
{Boquien}, M., {Duc}, P.-A., {Wu}, Y., {et~al.} 2009, \aj, 137, 4561

\bibitem[{{Boquien} {et~al.}(2011){Boquien}, {Lisenfeld}, {Duc}, {Braine},
  {Bournaud}, {Brinks}, \& {Charmandaris}}]{boquien11}
{Boquien}, M., {Lisenfeld}, U., {Duc}, P.-A., {et~al.} 2011, \aap, 533, A19

\bibitem[{{Bournaud}(2010)}]{bournaud10}
{Bournaud}, F. 2010, in Astronomical Society of the Pacific Conference Series,
  Vol. 423, Galaxy Wars: Stellar Populations and Star Formation in Interacting
  Galaxies, ed. B.~{Smith}, J.~{Higdon}, S.~{Higdon}, \& N.~{Bastian}, 177

\bibitem[{{Bournaud} {et~al.}(2007){Bournaud}, {Duc}, {Brinks}, {Boquien},
  {Amram}, {Lisenfeld}, {Koribalski}, {Walter}, \& {Charmandaris}}]{bournaud07}
{Bournaud}, F., {Duc}, P.-A., {Brinks}, E., {et~al.} 2007, Science, 316, 1166

\bibitem[{{Braine} {et~al.}(2003){Braine}, {Davoust}, {Zhu}, {Lisenfeld},
  {Motch}, \& {Seaquist}}]{braine03}
{Braine}, J., {Davoust}, E., {Zhu}, M., {et~al.} 2003, \aap, 408, L13

\bibitem[{{Braine} {et~al.}(2001){Braine}, {Duc}, {Lisenfeld}, {Charmandaris},
  {Vallejo}, {Leon}, \& {Brinks}}]{braine01}
{Braine}, J., {Duc}, P.-A., {Lisenfeld}, U., {et~al.} 2001, \aap, 378, 51

\bibitem[{{Braine} {et~al.}(2004){Braine}, {Lisenfeld}, {Duc}, {Brinks},
  {Charmandaris}, \& {Leon}}]{braine04}
{Braine}, J., {Lisenfeld}, U., {Duc}, P.-A., {et~al.} 2004, \aap, 418, 419

\bibitem[{{Braine} {et~al.}(2000){Braine}, {Lisenfeld}, {Duc}, \&
  {Leon}}]{braine00}
{Braine}, J., {Lisenfeld}, U., {Duc}, P.-A., \& {Leon}, S. 2000, \nat, 404, 904

\bibitem[{{Calzetti}(2013)}]{calzetti13}
{Calzetti}, D. 2013, {Star Formation Rate Indicators}, ed.
  J.~{Falc{\'o}n-Barroso} \& J.~H. {Knapen}, 419

\bibitem[{{Calzetti} {et~al.}(2007){Calzetti}, {Kennicutt}, {Engelbracht},
  {Leitherer}, {Draine}, {Kewley}, {Moustakas}, {Sosey}, {Dale}, {Gordon},
  {Helou}, {Hollenbach}, {Armus}, {Bendo}, {Bot}, {Buckalew}, {Jarrett}, {Li},
  {Meyer}, {Murphy}, {Prescott}, {Regan}, {Rieke}, {Roussel}, {Sheth}, {Smith},
  {Thornley}, \& {Walter}}]{calzetti07}
{Calzetti}, D., {Kennicutt}, R.~C., {Engelbracht}, C.~W., {et~al.} 2007, \apj,
  666, 870

\bibitem[{{Cardelli} {et~al.}(1989){Cardelli}, {Clayton}, \&
  {Mathis}}]{cardelli89}
{Cardelli}, J.~A., {Clayton}, G.~C., \& {Mathis}, J.~S. 1989, \apj, 345, 245

\bibitem[{{Casoli} {et~al.}(1998){Casoli}, {Sauty}, {Gerin}, {Boselli},
  {Fouque}, {Braine}, {Gavazzi}, {Lequeux}, \& {Dickey}}]{casoli98}
{Casoli}, F., {Sauty}, S., {Gerin}, M., {et~al.} 1998, \aap, 331, 451

\bibitem[{{Chabrier}(2003)}]{chabrier03}
{Chabrier}, G. 2003, \pasp, 115, 763

\bibitem[{{Chung} {et~al.}(2009){Chung}, {van Gorkom}, {Kenney}, {Crowl}, \&
  {Vollmer}}]{chung09}
{Chung}, A., {van Gorkom}, J.~H., {Kenney}, J.~D.~P., {Crowl}, H., \&
  {Vollmer}, B. 2009, \aj, 138, 1741

\bibitem[{{Daddi} {et~al.}(2010){Daddi}, {Elbaz}, {Walter}, {Bournaud},
  {Salmi}, {Carilli}, {Dannerbauer}, {Dickinson}, {Monaco}, \&
  {Riechers}}]{daddi10}
{Daddi}, E., {Elbaz}, D., {Walter}, F., {et~al.} 2010, \apjl, 714, L118

\bibitem[{{Dale} {et~al.}(2014){Dale}, {Helou}, {Magdis}, {Armus},
  {D{\'{\i}}az-Santos}, \& {Shi}}]{dale14}
{Dale}, D.~A., {Helou}, G., {Magdis}, G.~E., {et~al.} 2014, \apj, 784, 83

\bibitem[{{Dib} {et~al.}(2011){Dib}, {Piau}, {Mohanty}, \& {Braine}}]{dib11}
{Dib}, S., {Piau}, L., {Mohanty}, S., \& {Braine}, J. 2011, \mnras, 415, 3439

\bibitem[{{Duc}(2012)}]{duc12}
{Duc}, P.-A. 2012, {Birth, Life and Survival of Tidal Dwarf Galaxies}, ed.
  P.~{Papaderos}, S.~{Recchi}, \& G.~{Hensler}, 305

\bibitem[{{Duc} {et~al.}(2007){Duc}, {Braine}, {Lisenfeld}, {Brinks}, \&
  {Boquien}}]{duc07}
{Duc}, P.-A., {Braine}, J., {Lisenfeld}, U., {Brinks}, E., \& {Boquien}, M.
  2007, \aap, 475, 187

\bibitem[{{Duc} {et~al.}(2000){Duc}, {Brinks}, {Springel}, {Pichardo},
  {Weilbacher}, \& {Mirabel}}]{duc00}
{Duc}, P.-A., {Brinks}, E., {Springel}, V., {et~al.} 2000, \aj, 120, 1238

\bibitem[{{Duc} {et~al.}(2015){Duc}, {Cuillandre}, {Karabal}, {Cappellari},
  {Alatalo}, {Blitz}, {Bournaud}, {Bureau}, {Crocker}, {Davies}, {Davis}, {de
  Zeeuw}, {Emsellem}, {Khochfar}, {Krajnovi{\'c}}, {Kuntschner}, {McDermid},
  {Michel-Dansac}, {Morganti}, {Naab}, {Oosterloo}, {Paudel}, {Sarzi}, {Scott},
  {Serra}, {Weijmans}, \& {Young}}]{duc15}
{Duc}, P.-A., {Cuillandre}, J.-C., {Karabal}, E., {et~al.} 2015, \mnras, 446,
  120

\bibitem[{{Elmegreen}(1997)}]{elmegreen97}
{Elmegreen}, B.~G. 1997, in Revista Mexicana de Astronomia y Astrofisica
  Conference Series, Vol.~6, Revista Mexicana de Astronomia y Astrofisica
  Conference Series, ed. J.~{Franco}, R.~{Terlevich}, \& A.~{Serrano}, 165

\bibitem[{{Elmegreen} \& {Hunter}(2015)}]{elmegreen15a}
{Elmegreen}, B.~G. \& {Hunter}, D.~A. 2015, \apj, 805, 145

\bibitem[{{Elmegreen} {et~al.}(2013){Elmegreen}, {Rubio}, {Hunter}, {Verdugo},
  {Brinks}, \& {Schruba}}]{elmegreen13}
{Elmegreen}, B.~G., {Rubio}, M., {Hunter}, D.~A., {et~al.} 2013, \nat, 495, 487

\bibitem[{{Engelbracht} {et~al.}(2007){Engelbracht}, {Blaylock}, {Su}, {Rho},
  {Rieke}, {Muzerolle}, {Padgett}, {Hines}, {Gordon}, {Fadda},
  {Noriega-Crespo}, {Kelly}, {Latter}, {Hinz}, {Misselt}, {Morrison},
  {Stansberry}, {Shupe}, {Stolovy}, {Wheaton}, {Young}, {Neugebauer},
  {Wachter}, {P{\'e}rez-Gonz{\'a}lez}, {Frayer}, \& {Marleau}}]{engelbracht07}
{Engelbracht}, C.~W., {Blaylock}, M., {Su}, K.~Y.~L., {et~al.} 2007, \pasp,
  119, 994

\bibitem[{{Fazio} {et~al.}(2004){Fazio}, {Hora}, {Allen}, {Ashby}, {Barmby},
  {Deutsch}, {Huang}, {Kleiner}, {Marengo}, {Megeath}, {Melnick}, {Pahre},
  {Patten}, {Polizotti}, {Smith}, {Taylor}, {Wang}, {Willner}, {Hoffmann},
  {Pipher}, {Forrest}, {McMurty}, {McCreight}, {McKelvey}, {McMurray}, {Koch},
  {Moseley}, {Arendt}, {Mentzell}, {Marx}, {Losch}, {Mayman}, {Eichhorn},
  {Krebs}, {Jhabvala}, {Gezari}, {Fixsen}, {Flores}, {Shakoorzadeh}, {Jungo},
  {Hakun}, {Workman}, {Karpati}, {Kichak}, {Whitley}, {Mann}, {Tollestrup},
  {Eisenhardt}, {Stern}, {Gorjian}, {Bhattacharya}, {Carey}, {Nelson},
  {Glaccum}, {Lacy}, {Lowrance}, {Laine}, {Reach}, {Stauffer}, {Surace},
  {Wilson}, {Wright}, {Hoffman}, {Domingo}, \& {Cohen}}]{fazio04}
{Fazio}, G.~G., {Hora}, J.~L., {Allen}, L.~E., {et~al.} 2004, \apjs, 154, 10

\bibitem[{{Ferrarese} {et~al.}(2012){Ferrarese}, {C{\^o}t{\'e}}, {Cuillandre},
  {Gwyn}, {Peng}, {MacArthur}, {Duc}, {Boselli}, {Mei}, {Erben}, {McConnachie},
  {Durrell}, {Mihos}, {Jord{\'a}n}, {Lan{\c c}on}, {Puzia}, {Emsellem},
  {Balogh}, {Blakeslee}, {van Waerbeke}, {Gavazzi}, {Vollmer}, {Kavelaars},
  {Woods}, {Ball}, {Boissier}, {Courteau}, {Ferriere}, {Gavazzi},
  {Hildebrandt}, {Hudelot}, {Huertas-Company}, {Liu}, {McLaughlin}, {Mellier},
  {Milkeraitis}, {Schade}, {Balkowski}, {Bournaud}, {Carlberg}, {Chapman},
  {Hoekstra}, {Peng}, {Sawicki}, {Simard}, {Taylor}, {Tully}, {van Driel},
  {Wilson}, {Burdullis}, {Mahoney}, \& {Manset}}]{ferrarese12}
{Ferrarese}, L., {C{\^o}t{\'e}}, P., {Cuillandre}, J.-C., {et~al.} 2012, \apjs,
  200, 4

\bibitem[{{Genzel} {et~al.}(2010){Genzel}, {Tacconi}, {Gracia-Carpio},
  {Sternberg}, {Cooper}, {Shapiro}, {Bolatto}, {Bouch{\'e}}, {Bournaud},
  {Burkert}, {Combes}, {Comerford}, {Cox}, {Davis}, {Schreiber},
  {Garcia-Burillo}, {Lutz}, {Naab}, {Neri}, {Omont}, {Shapley}, \&
  {Weiner}}]{genzel10}
{Genzel}, R., {Tacconi}, L.~J., {Gracia-Carpio}, J., {et~al.} 2010, \mnras,
  407, 2091

\bibitem[{{Gil de Paz} {et~al.}(2003){Gil de Paz}, {Madore}, \&
  {Pevunova}}]{depaz03}
{Gil de Paz}, A., {Madore}, B.~F., \& {Pevunova}, O. 2003, \apjs, 147, 29

\bibitem[{{Gratier} {et~al.}(2010{\natexlab{a}}){Gratier}, {Braine},
  {Rodriguez-Fernandez}, {Israel}, {Schuster}, {Brouillet}, \&
  {Gardan}}]{gratier10a}
{Gratier}, P., {Braine}, J., {Rodriguez-Fernandez}, N.~J., {et~al.}
  2010{\natexlab{a}}, \aap, 512, A68

\bibitem[{{Gratier} {et~al.}(2010{\natexlab{b}}){Gratier}, {Braine},
  {Rodriguez-Fernandez}, {Schuster}, {Kramer}, {Xilouris}, {Tabatabaei},
  {Henkel}, {Corbelli}, {Israel}, {van der Werf}, {Calzetti}, {Garcia-Burillo},
  {Sievers}, {Combes}, {Wiklind}, {Brouillet}, {Herpin}, {Bontemps}, {Aalto},
  {Koribalski}, {van der Tak}, {Wiedner}, {R{\"o}llig}, \&
  {Mookerjea}}]{gratier10b}
{Gratier}, P., {Braine}, J., {Rodriguez-Fernandez}, N.~J., {et~al.}
  2010{\natexlab{b}}, \aap, 522, A3

\bibitem[{{Guillard} {et~al.}(2012){Guillard}, {Boulanger}, {Pineau des
  For{\^e}ts}, {Falgarone}, {Gusdorf}, {Cluver}, {Appleton}, {Lisenfeld},
  {Duc}, {Ogle}, \& {Xu}}]{guillard12}
{Guillard}, P., {Boulanger}, F., {Pineau des For{\^e}ts}, G., {et~al.} 2012,
  \apj, 749, 158

\bibitem[{{Hunter} {et~al.}(1998){Hunter}, {Elmegreen}, \& {Baker}}]{hunter98}
{Hunter}, D.~A., {Elmegreen}, B.~G., \& {Baker}, A.~L. 1998, \apj, 493, 595

\bibitem[{{J{\'a}chym} {et~al.}(2014){J{\'a}chym}, {Combes}, {Cortese}, {Sun},
  \& {Kenney}}]{jachym14}
{J{\'a}chym}, P., {Combes}, F., {Cortese}, L., {Sun}, M., \& {Kenney}, J.~D.~P.
  2014, \apj, 792, 11

\bibitem[{{Kennicutt} \& {Evans}(2012)}]{kennicutt12}
{Kennicutt}, R.~C. \& {Evans}, N.~J. 2012, \araa, 50, 531

\bibitem[{{Kennicutt}(1998)}]{kennicutt98}
{Kennicutt}, Jr., R.~C. 1998, \apj, 498, 541

\bibitem[{{Kroupa}(2001)}]{kroupa01}
{Kroupa}, P. 2001, \mnras, 322, 231

\bibitem[{{Krumholz} {et~al.}(2012){Krumholz}, {Dekel}, \&
  {McKee}}]{krumholz12}
{Krumholz}, M.~R., {Dekel}, A., \& {McKee}, C.~F. 2012, \apj, 745, 69

\bibitem[{{Lee} {et~al.}(2003){Lee}, {McCall}, \& {Richer}}]{lee03}
{Lee}, H., {McCall}, M.~L., \& {Richer}, M.~G. 2003, \aj, 125, 2975

\bibitem[{{Lee} {et~al.}(2009){Lee}, {Gil de Paz}, {Tremonti}, {Kennicutt},
  {Salim}, {Bothwell}, {Calzetti}, {Dalcanton}, {Dale}, {Engelbracht}, {Funes},
  {Johnson}, {Sakai}, {Skillman}, {van Zee}, {Walter}, \& {Weisz}}]{lee09}
{Lee}, J.~C., {Gil de Paz}, A., {Tremonti}, C., {et~al.} 2009, \apj, 706, 599

\bibitem[{{Lelli} {et~al.}(2015){Lelli}, {Duc}, {Brinks}, {Bournaud},
  {McGaugh}, {Lisenfeld}, {Weilbacher}, {Boquien}, {Revaz}, {Braine},
  {Koribalski}, \& {Belles}}]{lelli15}
{Lelli}, F., {Duc}, P.-A., {Brinks}, E., {et~al.} 2015, ArXiv e-prints

\bibitem[{{Lelli} {et~al.}(2014){Lelli}, {Fraternali}, \&
  {Verheijen}}]{lelli14}
{Lelli}, F., {Fraternali}, F., \& {Verheijen}, M. 2014, \aap, 563, A27

\bibitem[{{Lelli} {et~al.}(2012{\natexlab{a}}){Lelli}, {Verheijen},
  {Fraternali}, \& {Sancisi}}]{lelli12a}
{Lelli}, F., {Verheijen}, M., {Fraternali}, F., \& {Sancisi}, R.
  2012{\natexlab{a}}, \aap, 537, A72

\bibitem[{{Lelli} {et~al.}(2012{\natexlab{b}}){Lelli}, {Verheijen},
  {Fraternali}, \& {Sancisi}}]{lelli12b}
{Lelli}, F., {Verheijen}, M., {Fraternali}, F., \& {Sancisi}, R.
  2012{\natexlab{b}}, \aap, 544, A145

\bibitem[{{Leroy} {et~al.}(2012){Leroy}, {Bigiel}, {de Blok}, {Boissier},
  {Bolatto}, {Brinks}, {Madore}, {Munoz-Mateos}, {Murphy}, {Sandstrom},
  {Schruba}, \& {Walter}}]{leroy12}
{Leroy}, A.~K., {Bigiel}, F., {de Blok}, W.~J.~G., {et~al.} 2012, \aj, 144, 3

\bibitem[{{Leroy} {et~al.}(2008){Leroy}, {Walter}, {Brinks}, {Bigiel}, {de
  Blok}, {Madore}, \& {Thornley}}]{leroy08}
{Leroy}, A.~K., {Walter}, F., {Brinks}, E., {et~al.} 2008, \aj, 136, 2782

\bibitem[{{Lisenfeld} {et~al.}(2009){Lisenfeld}, {Bournaud}, {Brinks}, \&
  {Duc}}]{lisenfeld09}
{Lisenfeld}, U., {Bournaud}, F., {Brinks}, E., \& {Duc}, P.-A. 2009, ArXiv
  e-prints

\bibitem[{{Lisenfeld} {et~al.}(2004){Lisenfeld}, {Braine}, {Duc}, {Brinks},
  {Charmandaris}, \& {Leon}}]{lisenfeld04}
{Lisenfeld}, U., {Braine}, J., {Duc}, P.-A., {et~al.} 2004, \aap, 426, 471

\bibitem[{{Lisenfeld} {et~al.}(2002{\natexlab{a}}){Lisenfeld}, {Braine}, {Duc},
  {Leon}, {Charmandaris}, \& {Brinks}}]{lisenfeld02a}
{Lisenfeld}, U., {Braine}, J., {Duc}, P.-A., {et~al.} 2002{\natexlab{a}}, \aap,
  394, 823

\bibitem[{{Lisenfeld} {et~al.}(2002{\natexlab{b}}){Lisenfeld}, {Braine},
  {Vallejo}, {Duc}, {Leon}, {Brinks}, \& {Charmandaris}}]{lisenfeld02b}
{Lisenfeld}, U., {Braine}, J., {Vallejo}, O., {et~al.} 2002{\natexlab{b}}, in
  Astronomical Society of the Pacific Conference Series, Vol. 285, Modes of
  Star Formation and the Origin of Field Populations, ed. E.~K. {Grebel} \&
  W.~{Brandner}, 406

\bibitem[{{Lisenfeld} {et~al.}(2008){Lisenfeld}, {Mundell}, {Schinnerer},
  {Appleton}, \& {Allsopp}}]{lisenfeld08}
{Lisenfeld}, U., {Mundell}, C.~G., {Schinnerer}, E., {Appleton}, P.~N., \&
  {Allsopp}, J. 2008, \apj, 685, 181

\bibitem[{{Martin} \& {Kennicutt}(2001)}]{martin01}
{Martin}, C.~L. \& {Kennicutt}, Jr., R.~C. 2001, \apj, 555, 301

\bibitem[{{Morrissey} {et~al.}(2007){Morrissey}, {Conrow}, {Barlow}, {Small},
  {Seibert}, {Wyder}, {Budav{\'a}ri}, {Arnouts}, {Friedman}, {Forster},
  {Martin}, {Neff}, {Schiminovich}, {Bianchi}, {Donas}, {Heckman}, {Lee},
  {Madore}, {Milliard}, {Rich}, {Szalay}, {Welsh}, \& {Yi}}]{morrissey07}
{Morrissey}, P., {Conrow}, T., {Barlow}, T.~A., {et~al.} 2007, \apjs, 173, 682

\bibitem[{{Ploeckinger} {et~al.}(2014){Ploeckinger}, {Hensler}, {Recchi},
  {Mitchell}, \& {Kroupa}}]{ploeckinger14}
{Ploeckinger}, S., {Hensler}, G., {Recchi}, S., {Mitchell}, N., \& {Kroupa}, P.
  2014, \mnras, 437, 3980

\bibitem[{{Ploeckinger} {et~al.}(2015){Ploeckinger}, {Recchi}, {Hensler}, \&
  {Kroupa}}]{ploeckinger15}
{Ploeckinger}, S., {Recchi}, S., {Hensler}, G., \& {Kroupa}, P. 2015, \mnras,
  447, 2512

\bibitem[{{Raichoor} {et~al.}(2011){Raichoor}, {Mei}, {Nakata}, {Stanford},
  {Holden}, {Rettura}, {Huertas-Company}, {Postman}, {Rosati}, {Blakeslee},
  {Demarco}, {Eisenhardt}, {Illingworth}, {Jee}, {Kodama}, {Tanaka}, \&
  {White}}]{raichoor11}
{Raichoor}, A., {Mei}, S., {Nakata}, F., {et~al.} 2011, \apj, 732, 12

\bibitem[{{Rieke} {et~al.}(2004){Rieke}, {Young}, {Engelbracht}, {Kelly},
  {Low}, {Haller}, {Beeman}, {Gordon}, {Stansberry}, {Misselt}, {Cadien},
  {Morrison}, {Rivlis}, {Latter}, {Noriega-Crespo}, {Padgett}, {Stapelfeldt},
  {Hines}, {Egami}, {Muzerolle}, {Alonso-Herrero}, {Blaylock}, {Dole}, {Hinz},
  {Le Floc'h}, {Papovich}, {P{\'e}rez-Gonz{\'a}lez}, {Smith}, {Su}, {Bennett},
  {Frayer}, {Henderson}, {Lu}, {Masci}, {Pesenson}, {Rebull}, {Rho}, {Keene},
  {Stolovy}, {Wachter}, {Wheaton}, {Werner}, \& {Richards}}]{rieke04}
{Rieke}, G.~H., {Young}, E.~T., {Engelbracht}, C.~W., {et~al.} 2004, \apjs,
  154, 25

\bibitem[{{Schiminovich} {et~al.}(2007){Schiminovich}, {Wyder}, {Martin},
  {Johnson}, {Salim}, {Seibert}, {Treyer}, {Budav{\'a}ri}, {Hoopes},
  {Zamojski}, {Barlow}, {Forster}, {Friedman}, {Morrissey}, {Neff}, {Small},
  {Bianchi}, {Donas}, {Heckman}, {Lee}, {Madore}, {Milliard}, {Rich}, {Szalay},
  {Welsh}, \& {Yi}}]{Schiminovich07}
{Schiminovich}, D., {Wyder}, T.~K., {Martin}, D.~C., {et~al.} 2007, \apjs, 173,
  315

\bibitem[{{Schlafly} \& {Finkbeiner}(2011)}]{schlafly11}
{Schlafly}, E.~F. \& {Finkbeiner}, D.~P. 2011, \apj, 737, 103

\bibitem[{{Schlegel} {et~al.}(1998){Schlegel}, {Finkbeiner}, \&
  {Davis}}]{schlegel98}
{Schlegel}, D.~J., {Finkbeiner}, D.~P., \& {Davis}, M. 1998, \apj, 500, 525

\bibitem[{{Schruba} {et~al.}(2011){Schruba}, {Leroy}, {Walter}, {Bigiel},
  {Brinks}, {de Blok}, {Dumas}, {Kramer}, {Rosolowsky}, {Sandstrom},
  {Schuster}, {Usero}, {Weiss}, \& {Wiesemeyer}}]{schruba11}
{Schruba}, A., {Leroy}, A.~K., {Walter}, F., {et~al.} 2011, \aj, 142, 37

\bibitem[{{Shetty} {et~al.}(2011){Shetty}, {Glover}, {Dullemond}, {Ostriker},
  {Harris}, \& {Klessen}}]{shetty11}
{Shetty}, R., {Glover}, S.~C., {Dullemond}, C.~P., {et~al.} 2011, \mnras, 415,
  3253

\bibitem[{{Silk}(1997)}]{silk97}
{Silk}, J. 1997, \apj, 481, 703

\bibitem[{{Swaters} {et~al.}(2009){Swaters}, {Sancisi}, {van Albada}, \& {van
  der Hulst}}]{swaters09}
{Swaters}, R.~A., {Sancisi}, R., {van Albada}, T.~S., \& {van der Hulst}, J.~M.
  2009, \aap, 493, 871

\bibitem[{{Tan}(2000)}]{tan00}
{Tan}, J.~C. 2000, \apj, 536, 173

\bibitem[{{Toomre}(1964)}]{toomre64}
{Toomre}, A. 1964, \apj, 139, 1217

\bibitem[{{Verdugo} {et~al.}(2015){Verdugo}, {Combes}, {Dasyra}, {Salom{\'e}},
  \& {Braine}}]{verdugo15}
{Verdugo}, C., {Combes}, F., {Dasyra}, K., {Salom{\'e}}, P., \& {Braine}, J.
  2015, \aap, 582, A6

\bibitem[{{Verley} {et~al.}(2009){Verley}, {Corbelli}, {Giovanardi}, \&
  {Hunt}}]{verley09}
{Verley}, S., {Corbelli}, E., {Giovanardi}, C., \& {Hunt}, L.~K. 2009, \aap,
  493, 453

\bibitem[{{Werner} {et~al.}(2004){Werner}, {Roellig}, {Low}, {Rieke}, {Rieke},
  {Hoffmann}, {Young}, {Houck}, {Brandl}, {Fazio}, {Hora}, {Gehrz}, {Helou},
  {Soifer}, {Stauffer}, {Keene}, {Eisenhardt}, {Gallagher}, {Gautier}, {Irace},
  {Lawrence}, {Simmons}, {Van Cleve}, {Jura}, {Wright}, \&
  {Cruikshank}}]{werner04}
{Werner}, M.~W., {Roellig}, T.~L., {Low}, F.~J., {et~al.} 2004, \apjs, 154, 1

\bibitem[{{Wyder} {et~al.}(2009){Wyder}, {Martin}, {Barlow}, {Foster},
  {Friedman}, {Morrissey}, {Neff}, {Neill}, {Schiminovich}, {Seibert},
  {Bianchi}, {Donas}, {Heckman}, {Lee}, {Madore}, {Milliard}, {Rich}, {Szalay},
  \& {Yi}}]{wyder09}
{Wyder}, T.~K., {Martin}, D.~C., {Barlow}, T.~A., {et~al.} 2009, \apj, 696,
  1834

\bibitem[{{Zhu} {et~al.}(2008){Zhu}, {Wu}, {Cao}, \& {Li}}]{zhu08}
{Zhu}, Y.-N., {Wu}, H., {Cao}, C., \& {Li}, H.-N. 2008, \apj, 686, 155

\end{thebibliography}

%%%%%%%%%%%%%%%%%%%%%%
%%%% Appendix
%%%%%%%%%%%%%%%%%%%%%%

\appendix
\section{Channel maps}

\begin{figure*}[ht!]
 \centering
 \includegraphics[width=13cm,angle=0]{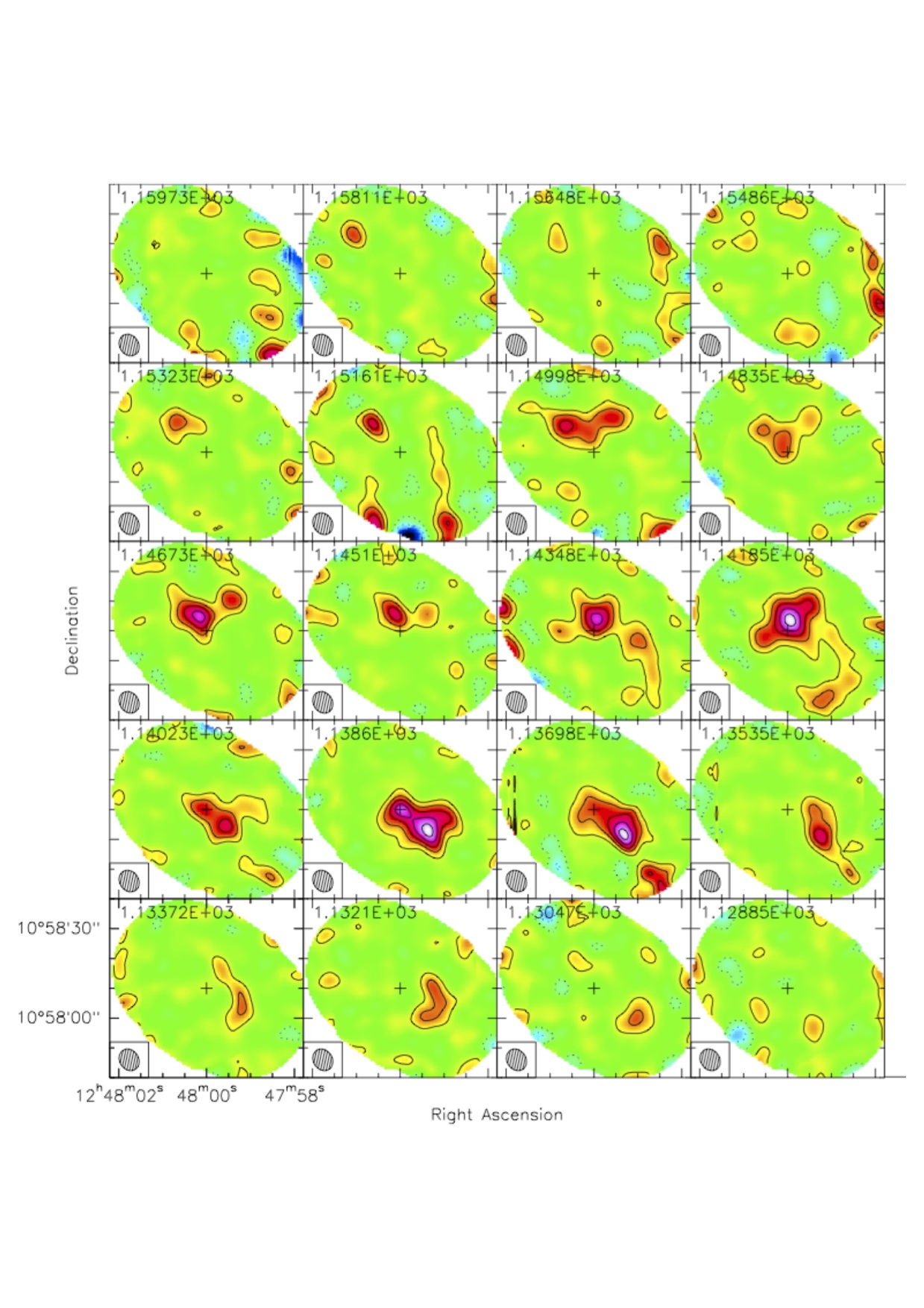}
 \caption{Channel maps of the emission tapered to 50m baseline length. The beam (size 7\farcs7 $\times$ 6\farcs3)
 is shown at the lower left corner of each image. The cross indicates the central position of the
 observations. 
 }
 \label{fig:channel-map}
\end{figure*}

Fig.~\ref{fig:channel-map} shows the channel maps of the taper50 data cube.
Emission is visible in the
velocity range between 1129 and 1159 \kms, distributed in various clouds and showing a velocity
gradient along the  NE-SW direction. 

\end{document}